\documentclass [smallcondensed, final]{svjour3}

\begin {document}

\sloppy

\hoffset 1cm \voffset -1cm %

\newcommand{\TITLE} { Matrix  theory of gravitation}
\markright{\large \TITLE \hfill \small W. K\"ohler \hspace*{1cm}}

\newcommand{\FN}[1]{\footnote { #1} } 
\newcommand{\MBOX}[1]{\quad \mbox {#1} \quad}
\newcommand{\HC}[1]{#1^\dagger} 

\newcommand{\MV}[1]{{\bf #1}} 
\newcommand{\TR}[1] {\mathcal T( #1 )} 
\newcommand{\VAR} {\delta \! } 
\newcommand{\LAGR} {\mathcal L} 
\newcommand{\lra} {\leftrightarrow} 
\newcommand{\BE}{\begin{equation}}
\newcommand{\EE}{\end{equation}}

\newcommand{\BEN}{\begin{eqnarray}} 
\newcommand{\EEN}{\end{eqnarray}}
\newcommand{\BEA}{\begin{eqnarray*}} 
\newcommand{\EEA}{\end{eqnarray*}}

\newcommand{\GT} {\mathcal G} 

\newcommand{\MATRIC}[1]  { \left( \begin{array}{ccccc} #1 \end{array} \right) }

\newcommand{\SC}[1]{{\sc #1}} 

\newcommand{\PD}[2][]{\frac{\partial #1}{\partial #2}} 
\newcommand{\DEF} {\stackrel {def} = }

\newcommand{\refeq}[1]{(\ref {#1})} 
\newcommand{\ORD}[2][] {\mathcal O (\epsilon^{#2} #1) }  
\newcommand{\VT} {\vartheta} 
\newcommand{\VP} {\varphi}

\title{ \TITLE\thanks{
    Published in ``General Relativity and Graviation'' (see {\tt http://springerlink.com/content/0001-7701/preprint/})\\
    The final publication is available at {\tt www.springerlink.com}
} }

\author{Wolfgang K\"ohler}
\institute{German Research Centre for Geosciences GFZ, D-14473 Potsdam,
  Germany  \hfill   \email{\tt wolfk@gfz-potsdam.de}\\   Homepage: {\tt  http://icgem.gfz-potsdam.de/Physics} 
}

\date {Received: 3 March 2010 / Accepted: 28 January 2011 / Published Online: 18 February 2011}

\maketitle

{\bf DOI: 10.1007/s10714-011-1158-x}\\

\begin{abstract}
  A new classical theory of gravitation within the framework of general relativity is presented.
  It is based on a matrix formulation of four-dimensional \SC{Riemann}-spaces and uses no 
  artificial fields or adjustable parameters. 
  The geometrical stress-energy tensor is derived from a matrix-trace \SC{Lagrangian}, which is not equivalent
  to the curvature scalar $R$. 
  To enable a direct comparison with the \SC{Einstein}-theory a tetrad formalism is utilized, 
  which shows similarities to teleparallel gravitation theories, but uses complex tetrads. Matrix theory might solve
  a 27-year-old, fundamental problem of those theories (Sect. \ref{sec-str-tetr}).
  For the standard test cases (PPN scheme, \SC{Schwarzschild}-solution)
  no differences to the  \SC{Einstein}-theory are found.   
  However, the matrix theory exhibits  novel, interesting vacuum solutions.
\end{abstract}

\setcounter{tocdepth}{2} 
{\small \tableofcontents}

\section{Introduction}
In the last decades a variety of new classical gravitation theories as alternatives to the \SC{Einstein}-theory 
were proposed \cite{WIKI}. This increased interest is particularly motivated by some new phenomena, which can only
be explained with some additional presumptions (e.g. galaxy rotation problem, Pioneer anomaly, accelerated Universe expansion).
On the other hand, new and enhanced
 experimental possibilities allow, to test their predictions \cite{WILL,FLECHTNER,ACES} with unthought precision.
We want to mention here only as representatives the Brans-Dicke theory \cite{BRANS}, as famous example of a scalar-tensor 
theory and MOND \cite{BEKENSTEIN}, which is supposed to
give an alternative to ``dark matter''. A recent discussion of this can be found in \cite{BROWST-MOFFAT}.
 
In this paper a new general relativistic gravitation theory, titled ``matrix theory'', is presented.
It is derived from a matrix-trace \SC{Lagrangian}, similar to the well-known \SC{Einstein}-\SC{Hilbert} action, 
but based on matrix formulation of the four-dimensional \SC{Riemannian} spacetime.\\
Like \SC{Einstein}'s original theory (without  the ``cosmological constant'') it contains no free, ``adjustable'' 
parameters, except the \SC{Newtonian} constant of gravitation $G$.
Also, it does not introduce new, artificial fields, like Brans-Dickes scalar-field or others in vector-tensor theories.

To compare it with the \SC{Einstein}-theory of gravitation, we generalize this  \SC{Lagrangian} with tetrad formalism,
so that it contains four real, constant parameters  $(a,b,c,d)$. Each parameter set 
then characterizes a different gravitation
theory, and it is shown that also the  \SC{Einstein}-theory belongs to this class of theories, esp. it is described
by the parameters $(a,b,c,d) = (1,-\frac 12, -\frac 14,0)$, while the matrix-theory is defined with $(1,-1,0,-\frac 12)$.

Matrix theory uses {\em complex} tetrads, because general base matrices $\tau_\mu$ can only be represented with such 
tetrads. This might look unfamiliar to some readers, but we consider this similar to the situation in quantum mechanics.
There we have a complex (non-measurable) wave function and real observables. Here, the tetrads themselves are also not
measurable, only the - by definition - real metric is measurable. Moreover, it shows, that 
all test cases computed here (sec. \ref{sec-compare}), which represent macroscopic matter 
(real, symmetric stress-energy tensor), have solutions with real tetrads
(for the PPN-test in section \ref{ssec-ppn} this holds up to the requested approximation order).

This tetrad formalism  shows, that the matrix theory can be regarded as generalization of the "teleparallel" 
approach (also called "distant parallelism" or 
"absolute parallelism") of tetrad gravity. 
This is based on an idea of Einstein,
which uses a non-symmetric ``Weitzenb\"ock'' connection with vanishing curvature
tensor but nonvanishing torsion, which is extensively discussed until 
today \cite{SHIRAFUJI,TUNG,PEREIRA,MALUF,HAMMOND}. A comprehensive overview can be 
found in \cite{HEHL-1998} and also \cite{HEHL-1996}, where the gauge aspects of the theory are stressed. 
If we would consider only real tetrads, the resulting theory (``RMT'', see sec. \ref{sec-RMT}) would belong to the one-parameter class 
of teleparallel theories, which are experimentally viable \cite{NESTER,HEHL-1998}, like the teleparallel equivalent of 
Einstein's GR (TEGR) \cite{MOLLER}. 
The teleparallel theory also allows an alternative {coframe} representation, which is used in \cite{ITIN-1,ITIN-2} to
derive a conserved energy-momentum current, completely similar to the Maxwell-Yang-Mills theory.\\
 However, the usage of complex tetrads,
which are necessary to map arbitrary matrices, and the new matrix-trace \SC{Lagrangian}, containing a parity violating term, exclude
typical ``unphysical''  tetrad vacuum solutions, which prevent a profound interpretation of previous teleparallel
theories.

In this paper, we use the conventional \SC{Levi-Civita}- (or \SC{Christoffel}-) connection, which is built from the metric 
tensor (see eq. \refeq{eq-tetr-cov-der} ff.).
The tetrad formalism here serves only as a general mathematical tool to compare different theories.
Instead of the tetrads, we consider the base matrices $\tau_\mu$ as the fundamental entities.  
Matter influence to geometry (field equations) is mediated via these $\tau_\mu$, resp. $\rho_{\mu\nu}$,  
but matter reacts to geometry only due to metric (equations of motion). 
Consequently, the geometrical stress-energy tensor is {\em potentially} not symmetric and real, 
but it is forced to be so, since it is equal to matter tensor (this is sometimes discussed differently for the teleparallel theory, 
see e.g. \cite{SHIRAFUJI}, p. 15).

However, many of the general tetrad computations presented here, are mostly standard (esp. the representation of
the \SC{Ricci}-scalar by tetrads and the derived stress-energy tensor)
and can be found at various places and in various contexts. E.g. our equation  \refeq{eq-EH-Lagr-tetr} is equivalent to 
eq. (1) in \cite{MALUF}.
The reason to sketch them here nevertheless, is to give a homogenous presentation with consistent notations.
This allows the reader to follow them without the necessity to check several sources with different names
for the same variables.

The readers will surely notice, that the hermitian matrices introduced in eq. \refeq{eq-mv-a} can also be 
regarded as second order \SC{Weyl}-spinors (see e.g. \cite{LL}, p. 59 ff)
with respect to their \SC{Lorentz}-transformation rule.
However, we do not discuss quantum mechanical effects or quantum fields here, to limit the extent of the paper.


\section{Formulation of the matrix theory}
To give a clearer picture, we start in \ref{sec-matr-gr} with the matrix representation of
the widely known {\em tetrad}- (or "Vierbein"-) formalism of general relativity 
(e.g. \cite{MTW}, \cite{STEPHANI}), which is given by {\em hermitian matrices} (real tetrads), 
and generalize this in section \ref{sec-general-matr} to general complex matrices (complex tetrads). 

\subsection{Hermitian matrix representation of tetrad formalism in relativity}
\label{sec-matr-gr}
Here we want shortly sketch, how the main tensors and equations of general and special relativity can be 
represented with hermitian matrices. This representation does not offer new equations, but it needs
less independent prerequisits (metric signature, {\sc Maxwell} eq.), than the usual component formulation.  
As far as we know, this cannot be found in the literature in this compact form.\\

{\bf Tetrads} are four real, covariant spacetime vectors, which are defined in each point of the spacetime. 
We denote them here by $e^a_\mu(x^\nu)$, where $a = 0,1,2,3$ is the tetrad index and
$\mu = 0,1,2,3$ the spacetime index (in this paper Greek letters 
$\mu,\nu,\alpha,\beta,\dots$ are used for
spacetime and Latin letters $a,b,c,d,\dots$ for tetrad indices). 
One of the first physicists, who used them  for GR (titled as ``four-legs'') was M{\o}ller, see e.g. his 
basic paper  \cite{MOLLER}.
As many others, he regards them as the fundamental gravitational field variables, instead of the metric $g_{\mu\nu}$.
Moreover, tetrads are also a useful tool in geodesic 
applications of general relativistic problems \cite{KUSCHE}.

Each individual tetrad (denoted by a certain fixed "$a$") is a  covariant tensor of first rank.
When $\eta_{ab} = diag[1,-1,-1,-1]$ denotes the \SC{Minkowski}-metric, the metric tensor $ g_{\mu\nu} $ is
expressed as
\BE \label{eq-def-g} 
g_{\mu\nu} = \eta_{ab} e_\mu^a e_\nu^b.
\EE
Tetrad indices $a,b,\dots$ can be shifted with $\eta_{ab}$ and $ \eta^{ab}$ while 
spacetime indices $\mu,\nu,\dots$ are shifted with $ g_{\mu\nu}$ and $ g^{\mu\nu}$.

The contravariant (inverse) tetrads  $e_a^\mu$ then fulfil two orthogonality relations, which are ($\delta$ is the usual \SC{Kronecker}-symbol):
\BE \label{eq-tetr-orthogon}
e^\mu_a e^a_\nu = \delta^\mu_\nu\MBOX{and}  e^\mu_a e^b_\mu = \delta^b_a.
\EE
By contracting with $e_a^\mu $ or $e^a_\mu$
any spacetime index of any symbol (tensor or non-covariant entity), can be transformed into a tetrad index, and  
vice versa, e.g.
\BE A^\mu e_\mu^a = A^a \quad \lra \quad A^\mu = A^a e_a^\mu.
\EE
{\bf Matrix representation}:
with the tetrads one can construct four complex, hermitian $2\times 2$-matrices, using the
generalized \SC{Pauli} spin matrices $\sigma_0 =  {1 0\choose 0 1} = \MV I_2,\; \sigma_1 = {0 1 \choose 1 0},\;
\sigma_2 = {0 -i\choose i\;\; 0},\; \sigma_3 = {1 \;\; 0\choose 0 -1}$, 
which we will denote with $\tau_\mu$:\FN{
  More generally, any set of 4 hermitian matrices $\sigma_m'$ can be used as basis, that preserves the orthogonality
  $\frac 12 \TR{\sigma_m' \bar\sigma_n'} = \eta_{mn}$. This is in close relation to the transformations
  described in eq. \refeq{eq-T-trafo}.
}
\BE \label{eq-def-tau} 
 \underline{ \tau_\mu \DEF e_\mu^a\sigma_a}.
\EE
This definition is very similar to the expression of spinor components of tensors with the help of
Infeld - van der Waerden symbols (\cite{PENROSE}, p. 123 and \cite{STEPHANI}, p. 48) 
$g_a^{A B'} = \frac 1{\sqrt 2}\sigma_a^{AB'}$ where $A,B' \in [1,2] $ are the spinor indices.

These four matrices $\tau_\mu $ are hermitian by construction, linearly independent, and can replace 
the tetrads, since eq. \refeq{eq-def-tau} is an invertible map. 
They form a basis in the vector space of $2\times 2$ matrices, like the four $\sigma_k$.\FN{
  I.e. every (hermitian) matrix $\MV A$ can be expressed as linear combination $\MV A = a^\mu \tau_\mu$, with complex (real)
  coefficients $a^\mu$.
} 
We will denote them here as "base matrices". 
Some general relations with these matrices are listed in the appendix \ref{ssec-matrix}.

Any tensor of first rank with contravariant components $A^\mu$ can be expressed  
as hermitian matrix $\MV A$ by (boldface Latin letters $\MV A,\MV B, \dots$ as well as Greek letters $\tau,\sigma,\rho,\dots $ 
shall denote hermitian $2\times 2$-matrices here) 
\BE\label{eq-mv-a} 
  \MV A = A^\mu \tau_\mu.
\EE
Since the base matrices build a covariant ``tensor-matrix'', this means that
the matrix $\MV A$ is actually {\em invariant} under all transformations $x'^{\mu} (x^\nu)$.
Of course, also the infinitesimal line element (1-form) can be expressed as the matrix $\MV {dx} = dx^\mu \tau_\mu$ and 
transformation equations are
\BE \label{eq-coord-trafo}
  dx'^{\mu} = \PD[x'^{\mu}]{x^\nu} dx^\nu = a_\nu^{\mu} dx^\nu \MBOX{and} \tau_\nu 
  = a_\nu^{\mu} \tau'_{\mu} .
\EE
This transformation rule for the base matrices states, that all components are transformed
with the same coefficients (like the tetrads).\\
 
A novelty of the matrix notation, in contrast to usual tetrad notation, is, that it defines an {\em inner product} (matrix product) and with the 
help of this, the need to postulate a \SC{Minkowski} norm for the tetrads 
(with its sign-arbitrariness) disappears.  
This hermitian matrix-algebra can be seen as special representation of Heestenes' ``space-time algebra'' (STA), which is widely discussed
in the literature, especially for the \SC{Dirac}-theory, see e.g. \cite{GULL}.

The norm of a tensor $A_\mu$ is the simple matrix-determinant 
\BE \label{eq-matr-norm}
|\MV A| =  g_{\mu\nu} A^\mu A^\nu  =  A_\mu A^\mu.
\EE
This is easy to derive from the properties of the $\sigma_k$,  namely $\frac 12\TR{\sigma_m \bar \sigma_n} = \eta_{mn}$,
where $\TR{\MV A} $ denotes the trace and $\bar\MV A$ the ``adjuncted'' matrix\FN{
  This term is not widely used in English mathematical textbooks. We define it here as $\bar\MV A \DEF |\MV A|  \MV A^{-1}$.
  Please note, that  only for $2\times 2$ matrices we have the {\em linear map}  $\MV A \lra \bar \MV A$
  and consequently only then $|\MV A| = \frac 12\TR{\MV A \bar\MV A} $ is a {\em bilinear form}.
  See appendix \ref{ssec-matrix} for a more detailed discussion.
}  of a matrix $\MV A$.
This simple norm definition is only possible for a four-dimensional \SC{Riemannian} spacetime with 
\SC{Minkowskian} signature $[+,-,-,-]$.

Additionally we have to introduce the contravariant basis $\tau^\mu = g^{\mu\nu} \tau_\nu$ and can
derive the orthogonality relations 
\BE \label{eq-g-matr-def}
 \frac 12\TR{\tau_\mu \bar \tau_\nu} = g_{\mu\nu} \MBOX{and} \frac 12 \TR{\tau_\mu \bar \tau^\nu} = \delta_\mu^\nu.
\EE
If the matrix theory is formulated without tetrads, the first equation 
is to interpret as the definition of the metric tensor $ g_{\mu\nu}$ and the second as the definition of 
the matrices $\tau^\mu$.

The more general scalar product of two tensors $\MV A, \MV B$ has a similar matrix representation
like the norm in eq. \refeq{eq-matr-norm}
\BE \frac 12 \TR{\MV A\bar \MV B} = g_{\mu\nu} A^\mu B^\nu = A_\mu B^\mu.
\EE
The inverse relation of eq. \refeq{eq-mv-a} is the trace expression (always real)
\BE A^\mu = \frac 12 \TR{\MV A \bar\tau^\mu} \MBOX{and}  A_\mu = \frac 12 \TR{\MV A \bar\tau_\mu}.
\EE
Tensors of higher rank are expressed by sets of hermitian matrices, e.g. a general tensor of second rank 
with four matrices
\BE \MV A_\mu = A_{\mu\nu} \tau^\nu \quad \lra \quad A_{\mu\nu} = \frac 12\TR{\MV A_\mu\bar \tau_\nu}.
\EE
With the above definitions the complete apparatus of special and general relativity can be drawn in matrix form. 
E.g. the covariant derivative of the basis is computed like for a conventional vector
\BE \tau_{\mu;\nu} \DEF \tau_{\mu,\nu} - \Gamma^\lambda_{\mu\nu} \tau_\lambda.
\EE 
The \SC{Christoffel} symbols  $\Gamma^\lambda_{\mu\nu}$ 
defining the connection here, have to be derived metric compatible from eq. \refeq{eq-def-g} (see appendix \ref{ssec-append-1}). 
The matrix-representation of the antisymmetric second
covariant derivatives of the basis then gives a definition of a \SC{Riemann} tensor matrix, which is very similar to the standard formula:
\BE \tau_{\mu;\nu;\lambda} -\tau_{\mu;\lambda;\nu} = R^\sigma_{\;\;\mu\nu\lambda} \tau_\sigma 
\DEF  \MV R_{\mu\nu\lambda}.
\EE
Interested readers can also have a look at \cite{KOE-Mink}, where representations of
main topics of  {\em special relativity} (e.g. electromagnetism, \SC{Dirac}-equation)
with matrices are shown.
As single example we cite here the matrix representation of {\sc Maxwells} equations
($\partial \DEF \sigma_\mu \frac {\partial}{\partial x^\mu}$ is the partial derivation operator matrix, $\MV F = (E_k + i B_k)\sigma^k$ 
the trace-free electromagnetic field matrix (non-herm.) and $\MV J$ is the hermit. current matrix):\FN{
  In flat {\sc Minkowski} spacetime we use $\tau_\mu \equiv \sigma_\mu = const.$
}
\BE \partial \MV F = \MV J.
\EE
This is only one matrix eq., 
but it contains 8 real (4 complex) component equations, which are the four homogeneous (anti-hermit. part)
and the four inhomogeneous Maxwell eqs. (hermit. part).\\

Additionally to local spacetime covariance, the matrix equations exhibit another, independent {\em global symmetry}:
If all matrices are synchronously transformed with one 
constant, unimodular matrix $\MV T$ (i.e. $|\MV T| = 1$), preserving their hermitian property:
\BE \label{eq-T-trafo}
\MV A \to \MV T \MV A \HC {\MV T},
\EE
then obviously all relations, e.g. the metric in eq. \refeq{eq-g-matr-def}, remain unchanged.
The transformation matrix $\MV T$ then contains $6$ real parameters 
and it is easy to show, that it can be identified with a \SC{Lorentz}-transformation in a local 
\SC{Minkowski}-coordinate system.\FN{
  The group of matrices $\MV T$ with complex elements, satisfying $|\MV T| = 1$, is commonly denoted as $SL(2, \mathcal C)$. 
  It is the "double cover" of the Lorentz-group, because both matrices
  $\MV T$ and $(- \MV T)$ perform the same Minkowski-space rotation.
} 
Consequently, in those coordinate systems (locally) both transformations may be combined arbitrarily. \\

To describe curvature in \SC{Riemannian} geometry, we define 
the "rho"-tensor-matrix, as the antisymmetric partial derivative of the basis
\BE \label{eq-def-rho}
   \underline{\rho_{\mu\nu} \DEF \tau_{\mu,\nu} - \tau_{\nu,\mu}}.
\EE
The tensor property (covariant transformation rule) of this matrix-tensor is evident. 
It consists of 6 hermitian matrices and thus 
contains $4\times 6 = 24$ real components.
From  $\rho_{\mu\nu} \equiv 0$ follows
the vanishing of the \SC{Riemann}-tensor $R^\sigma_{\;\;\mu\nu\lambda} = 0$ (e.g. by the derivations in 
the appendix \ref{ssec-append-1}) i.e. the spacetime is flat.
On the other hand, for a flat spacetime we can always find a coordinate system with $\tau_\mu = const.$ and
consequently $\rho_{\mu\nu} = 0$. Due to the tensor property this equation remains true, if an arbitrary
coordinate transformation is applied.

From the definition of $\rho_{\mu\nu}$ and the basis in eq. \refeq{eq-def-tau} we get the tetrad formula
(we use the common $[]$-bracket-notation, but omit a frequently used factor $1/2$)
\BE \label{eq-rho-e}
  \rho_{\alpha\gamma} =  (e^x_{\alpha,\gamma} - e^x_{\gamma,\alpha})\sigma_x \DEF e^x_{[\alpha,\gamma]}\sigma_x,
\EE 
where $ e^x_{[\alpha,\gamma]}$ is the "nonholonomity" \cite{HAMMOND}.
Since the \SC{Pauli} matrices are constant, it is evident that  $\rho_{\mu\nu}$ is the matrix representation of the 
exterior derivative of the basis 1-forms $\theta^x =  e^x_{\mu} dx^\mu $. 

The expressions are further simplified by transforming the  spacetime indices $\alpha,\gamma$ 
into tetrad-indices.  For this purpose, we define the antisymmetric tetrad expressions $r^x_{ac} = -r^x_{ca}$ by
(Schouten, \cite{SCHOUTEN}, pp. 99, denotes them as ``objects of anholonomy'' $\Omega^x_{ac}$):
\BE\label{eq-def-rafb} 
  r^x_{ac} \DEF e^x_{[\alpha,\gamma]} e^\alpha_a e^\gamma_c.
\EE
\label{r-types}
These 24 coefficients $ r^x_{ac}$ can be classified into {\em two types}. 
For 12 of them the upper index $x$ is equal to one of the lower. They will be denoted here as "r-doublets". 
The other 12, where all three indices $x\ne a\ne c$ are different, are denoted as "r-triplets".
This classification is independent of the
coordinate system, because the  $ r^x_{ac}$ are invariant under all coordinate transformations.

By Cartan's first structural equations one can see, that these terms are closely related  to the
``Ricci rotation coefficients'', which may be defined from the covariant derivative (also see appendix \ref{ssec-append-1})
\BE   \GT^s_{mn}  \DEF  e^\mu_m e^\nu_n e^s_{\mu; \nu} =
  \frac 12 (r^s_{mn} + \eta^{sb}(\eta_{mc} r^c_{nb} +\eta_{nc} r^c_{mb})) .
\EE
From these one can directly derive the tetrad representation of the curvature tensor (see eq. \refeq{eq-Riemann-def} ff.)
\BE 
    { R^s_{\;mnl}=  e^\lambda_p (\delta^p_l \GT^s_{mn} - \delta^p_n \GT^s_{ml} )_{,\lambda}
      + \GT^s_{xy}( \delta^x_m r^y_{nl} + \delta^y_n \GT^x_{ml} - \delta^y_l\GT^x_{mn})}.
\EE
\subsection{General matrices and complex tetrads}
\label{sec-general-matr}
For the matrix representation presented above, it looks straightforward, to consider general 
instead of special (hermitian) matrices as basis $\tau_\mu$. 
Another motivation comes from quantum mechanics, which cannot be formulated without complex wave functions.  
Therefore one may hope, that the ideas presented here can help to find a new link between quantum mechanics and gravity.
However, this is not the topic of this paper, which covers only classical gravity.\\
On the other hand, for tetrad gravity in usual formulation, it makes no sense to introduce complex - instead 
of real - tetrads, because the field equations are not altered. 
This is only the case, if we use the matrix-{\sc Lagrangian}
defined in \ref{sec-lagr-matr}, which has additional complex terms.\\

For general base matrices, we have to generalize the metric definition in eq. \refeq{eq-g-matr-def}, because 
the distance  $ds^2 = g_{\mu\nu} dx^\mu dx^\nu$ must always be a real quantity.\\ 
Regarding that under coordinate transformations the hermitian conjugated matrices $ \HC \tau_\mu $ obey the same 
transformation rule as $\tau_\mu$ in eq.
\refeq{eq-coord-trafo} (the transformation coefficients $a^\mu_\nu$ are real),
the appropriate definition is\FN{
  One might also discuss to use the complex value (without $\Re$), which would define a hermitian metric tensor
  $g_{\mu\nu}^\ast = g_{\nu\mu}$. For all equations, where only the symmetric part of $ g_{\nu\mu}$ occurs, e.g. the 
  equation of motion, it is equivalent. This form also allows the usual index shifting with $g$.
}
\BE 
  \label{eq-def-metrik}
  g_{\mu\nu} \DEF \frac 14 \TR{\HC\tau_\mu\bar\tau_\nu +\HC\tau_\nu\bar\tau_\mu } =  \frac 12 \Re\TR{\HC\tau_\mu\bar\tau_\nu },  
\EE
which is symmetric and real for arbitrary matrices $\tau_\mu$ and for hermitians $\tau_\mu = \HC \tau_\mu $ it is equal 
to the definition in \refeq{eq-g-matr-def}.
It formally resembles definitions of quantum mechanical observables, e.g. the {\sc Dirac} current and is invariant under 
unitary $U(1)$ (phase) transformations $\tau_\mu \to e^{i\VP} \tau_\mu$, additionally to its $T$-invariance described in \refeq{eq-T-trafo}.

 Also the scalar product of two tensor matrices
$\MV A = a^\mu \tau_\mu,\; \MV B = b^\mu \tau_\mu$ (with $a^\mu, b^\mu = real$)
is to define consistently as the real number
\BE
  (\MV A \cdot \MV B) \DEF  \frac 12\Re \TR{\MV {\HC A} \MV {\bar B}}
  = a^\mu b^\nu \frac 12\Re  \TR{\HC\tau_\mu\bar\tau_\nu } = a^\mu b^\nu g_{\mu\nu}.
\EE
With this definition all equations of general relativity stay valid, except the field equations.
For clarity we add, that for all physical problems covered here, 
we consider only strictly {\em real and symmetric} stress-energy tensors $T^{\mu\nu}$ of matter.
This requires, that all possible imaginary and anti-symmetric parts of the geometric tensor also vanish. 
We do not discuss possible implications of
those terms, instead we demand that all must be zero, for all classical gravity problems in this paper. Then, e.g. for the test cases
solved with real tetrads in section \ref{sec-compare}, the imaginary parts form  additional constraints, compared
to a  corresponding real-tetrad theory (``RMT'', see sec. \ref{sec-RMT}).
 
For the definition of the contravariant base matrices $\tau^\mu$ we cannot use the metric $g^{\mu\nu}$ here anymore, 
but the second eq. of \refeq{eq-g-matr-def} gives an unique definition. 
The contravariant transformation rule stays valid, due to this orthogonality relations.

If we want to utilize the tetrad formalism for general bases, we have to use {\em complex tetrads} in the 
decomposition $\tau_\mu = e^a_\mu \sigma_a$. 
The inverse tetrads $e^\mu_a$
are also to define with their orthogonality relations in eq. \refeq{eq-tetr-orthogon}, index shifting with $g^{\mu\nu}$ 
is also not applicable for them.

We have to add here, that for general (non-hermitian) matrices $\tau_\mu$, 
the metric is not necessarily locally Lorentzian. However, this is always true for the physically important
case, when the imaginary parts of the tetrads are small (e.g. for the PPN-tests in sec. \ref{ssec-ppn}).
General matrices $\tau_\mu$ can be decomposed into a hermitian and anti-hermitian part, 
and correspondingly the complex tetrads $e^a_\mu$ into real and imaginary parts:
$e^a_\mu = f^a_\mu + ih^a_\mu $ (with $f^a_\mu, h^a_\mu = real$). The metric definition \refeq{eq-def-metrik} then gives
$g_{\mu\nu} = (f^a_\mu f^b_\nu + h^a_\mu h^b_\nu )\eta_{ab} $.\\ 
It can be shown, that it is locally Lorentzian, if all imaginary parts are small: $||h^a_\mu|| \ll 1,\; \forall a, \mu $.

\subsection{\SC{Lagrangian} of matrix-theory}
\label{sec-lagr-matr}
It is an important feature of the \SC{Einstein}-equation in general relativity, that it can be derived from
a \SC{Lagrangian} $\LAGR$ (see e.g. \cite{STEPHANI}), 
namely its geometrical part equals the curvature scalar $\LAGR_E \simeq R$.\\ 
For deriving the stress-energy tensor and the field equations one has to 
find the stationary solution of the action integral
\BE \label{eq-action-einst}  
  I = \int d^4 x \sqrt{-g} \LAGR(g_{\mu\nu},\; g_{\mu\nu,\lambda})
\EE
by variation of the metric tensor $\VAR g_{\mu\nu}$.
The same holds for the
matrix theory, where we postulate another scalar based on the "rho"-tensor-matrix defined in eq. \refeq{eq-def-rho}.
As explained, this tensor-matrix also characterizes the curvature of spacetime and it is straight forward
to construct a theory  of gravity based on this tensor-matrix.

Here we construct the ``matrix-\SC{Lagrangian}'' $\LAGR_z$ as real, scalar, bilinear form from the matrices 
$\rho_{\mu\nu}$ and $ \HC\rho_{\mu\nu}$. We request the same symmetry as for the metric definition \refeq{eq-def-metrik}, 
i.e. global $T$-invariance forces, that the matrix factors in the trace must build a bar-alternating sequence.

There exist only two distinct tensor matrices, that can be built by bar-alternating contraction of $\rho_{\mu\nu}$, 
namely $ \rho_\mu \DEF  \bar\tau^\nu\rho_{\mu\nu} $ and $ \upsilon_\mu \DEF \rho_{\mu\nu}\bar\tau^\nu$. With the request of 
unitary $U(1)$ invariance we postulate the following  \SC{Lagrangian}, which is also quadratic in the first derivatives:\FN{
  The more general \SC{Lagrangian} of ``viable'' theories $\LAGR_v$, which is discussed in section \ref{sec-iso-sm} for comparison,
  can be written in the same form, with an extra term (exhibiting the same symmetries, but real by definition): 
  $ \LAGR_v(c) = \LAGR_z +  \frac c4 \TR{\tau^{\alpha\dagger} \bar \tau^\beta \upsilon_{\beta} \bar\rho_{\alpha}^\dagger }$,
  where ``$c$'' is a free, real constant. With the terminology of section \ref{sec-gen-lagr} for the extra term holds 
  $ \frac 14 \TR{\tau^{\alpha\dagger} \bar \tau^\beta \upsilon_{\beta}\bar\rho_{\alpha}^\dagger } = \LAGR_c - 2\LAGR_b$.
}
\BE \label{eq-def-lagr-matr}
\fbox{ $
  \LAGR_z \DEF \frac 14 \Re\TR{\tau^{\alpha\dagger} \bar \tau^\beta \rho_{\alpha}^\dagger \upsilon_{\beta}}.
$}
\EE 
This expression is a real function of the $\tau_\mu, \HC\tau_\mu$ and their first derivatives 
$\LAGR_z(\tau_\mu,\HC \tau_\mu, \tau_{\mu,\nu},\HC\tau_{\mu,\nu})$ and contains { no adjustable parameters}.
By construction, it is invariant under arbitrary coordinate transformations 
and constant (global) $T$-transformations described in  eq. \refeq{eq-T-trafo}. 
\label{Lagr-Symm} Considering its additional unitary invariance  under
$\tau_\mu \to e^{i\VP} \tau_\mu$, we find that the symmetry group is $ SL(2, \mathcal C) \times U(1) $, which is a supergroup of  $SU(2) \times U(1)$,
the important group of standard electro-weak ``GSW-theory'' (see \cite{EBERT}, we only discuss global symmetry here).\\
 For completeness we have to add, that \refeq{eq-def-lagr-matr} is of course not the only possible form. In general, every bar-alternating permutation 
of the 6 factors $\tau^{\alpha\dagger}, \tau^{\gamma\dagger}, \rho_{\alpha\gamma}^\dagger, \tau^\beta, \tau^\delta, \rho_{\beta\delta}$ 
exhibits the same symmetries and its tetrad \SC{Lagrangian} has the common form \refeq{eq-Lag-abcde}. 
But if we request, that the contracted forms $ \rho_\mu $ and $ \upsilon_\mu$ should occur, 
but {\em no doubly contracted} matrices (like  $\bar\tau^\nu\rho_{\mu\nu} \bar\tau^\mu$), then only four alternatives remain:
$\Re\TR{\tau^{\alpha\dagger} \bar \tau^\beta \MV x_{\alpha\beta}}$, where 
$\MV x_{\alpha\beta} =  \rho_{\alpha}^\dagger \upsilon_\beta,\;  \bar \rho_{\alpha}^\dagger \upsilon_\beta,  \; 
\rho_{\alpha}^\dagger \bar \upsilon_\beta, \;   \bar \rho_{\alpha}^\dagger \bar \upsilon_\beta $, respectively. The fourth alternative gives a 
completely similar \SC{Lagrangian} as the first eq. \refeq{eq-def-lagr-matr}
(namely $\LAGR = \LAGR(1,-1,0,+\frac 12)$ in eq. \refeq{eq-Lag-abcde}, i.e. only the odd parity term $\LAGR_i$ has opposite sign, 
which does not affect any conclusions), while the second and third form have no odd parity term.

To derive the field eqs., similarly to above eq. \refeq{eq-action-einst}, one could vary the base matrices $\VAR \tau_\mu$ instead of $\VAR g_{\mu\nu}$\FN{
  Here is to replace $ 4 \sqrt{-g} = ||\tau||$, where  $||\tau||$ is defined as  absolute value of the determinant $|\tau|$ of 
  all $4\times 4$ components of the basis.
  For the variation of this term one has to use $\VAR|\tau| = \frac {|\tau|} 2 \TR{\VAR \tau_\mu \bar \tau^\mu}$
}
\BE
 I =  \int d^4x ||\tau|| \LAGR(\tau_\mu,\tau_{\mu,\nu}) \quad\to \quad \VAR I = \int d^4 x ||\tau|| \TR{\VAR \tau_\mu \bar \MV T^\mu}
\EE 
This derivation of the stress-energy tensor matrix $\MV T^\mu$ would be straightforward.
But instead of this, we give an equivalent derivation with the use of tetrads in the next paragraph. 
This has the advantage to be more general and so allows a direct comparison to the \SC{Einstein}-theory.\\

{\bf Tetrad representation} of $\LAGR_z$:\\
With the terms in eq. \refeq{eq-rho-e} the \SC{Lagrangian} in  \refeq{eq-def-lagr-matr} can be rewritten as (all $r^x_{ab}$
are scalar and can be drawn out of the matrix trace):
\BE  \label{eq-lagr-tetr}
  \LAGR_{z} = \frac 14 \Re\TR{\tau^{\alpha\dagger}\bar\tau^{\beta}
    \rho_{\alpha\gamma}^\dagger\bar \tau^{\gamma\dagger} \rho_{\beta\delta}  \bar\tau^{\delta }}
  = \frac 14 \Re \big( (r^{x}_{ac})^* r^y_{bd} \TR{\sigma^a\bar\sigma^b \sigma_x \bar \sigma^c \sigma_y\bar\sigma^d  } \big)
\EE
The trace of 6 {\sc Pauli}-matrices above is computed using the techniques in the appendix \ref{ssec-matrix}. 
If we define for abbreviation the two contracted terms ($r_a$ is constructed on only of r-doublets and $t^a$ only of r-triplets): 
\BE \label{eq-def-rt}
  r_a \DEF r^x_{ax} \MBOX {and} t^a \DEF \frac 12 \eta_{y b} r^{y}_{c d} \Delta^{ a b c d},
\EE 
where $\Delta^{abcd}$ is the completely antisymmetric symbol, with  $\Delta^{0123} = 1$, the result is
\BE \label{eq-lagr_z}
  \LAGR_z = \underbrace{ \eta^{m n} (r_{m} r^{\ast}_{n} -r^{a}_{m b} (r^{b}_{na})^*) }_{\DEF \LAGR_r}
  + i\underbrace{ ( t^a r^{\ast}_{a} - t^{a*} r_{a}) }_{\DEF \LAGR_i}.
\EE 
Both terms $\LAGR_r $ and $ i \LAGR_i$ are evidently real ($\LAGR_i^\ast = -\LAGR_i$).\\
The explicit appearance of the imaginary unit ``$i$'' in this formula is a consequence of utilizing
{\sc Pauli}-matrices $\sigma_a$ as basis in $\tau_\mu = e^a_\mu \sigma_a$.\FN{
 They are well suited as basis for hermitian matrices, but not 
 the best choice for arbitrary complex matrices.
  Another choice are the matrix-components themselves, which 
  leads to a spinor-like notation $\tau_\mu = t^{AB}_\mu\VT_{AB}$, $ A,B = (1,2) $,
  with $\VT_{11} \DEF {10 \choose 00}, \quad \VT_{12} \DEF {01 \choose 00},\; \dots $.
  The {\sc Lagrangian}, computed with these 16 complex terms $t^{AB}_\mu$, instead of the tetrads $e^a_\mu$,
  is somewhat simpler. However, they must be transformed into tetrads anyway, 
  to describe local {\sc Minkowski}-systems and for the test cases of section \ref{sec-compare}.
}\\
We note, that $\LAGR_i$ has ``odd parity'' (due to the factor $\Delta$) 
in contrast to all other terms of $\LAGR_r$ and eq. \refeq{eq-Lag-abcde}) with respect to the tetrad space 
(the tetrad parity operation is equivalent to the matrix transformation $\tau^\mu \to \bar \tau^\mu$, which inverts 
the three spatial tetrads $e^k_\mu, k= 1,2,3$). 

\label{disc-rc-var}
We now discuss the implications of using {\em real or complex} tetrads for the variation principle.
The variation gives the definition of the stress-energy tensor components $ T^\gamma_h$ by
\BE 
  \int \VAR\LAGR_z = \int  \VAR\LAGR_r + i \VAR \LAGR_i \DEF  
  \int \VAR e^h_\gamma T^\gamma_h + (\VAR e^h_\gamma)^\ast (T^\gamma_h)^\ast \stackrel != 0. 
\EE
If we consider only {\em a priori  real tetrads}, like in conventional tetrad theories, also the 
variations must be real, i.e. $(\VAR e^h_\gamma)^\ast  = \VAR e^h_\gamma $, and the variation principle gives only
$ (T^\gamma_h)^\ast + T^\gamma_h = 2\Re( T^\gamma_h) = 0$, which means that $\LAGR_i$ does not contribute in this case
(the resulting theory ``RMT'' is discussed in sec. \ref{sec-RMT}).\\ 
For the case of {\em potentially complex tetrads}, both variations 
 $\VAR e^h_\gamma ,\; (\VAR e^h_\gamma)^\ast $ are independent and we get the full complex eq. $T^\gamma_h = 0$ and both 
$\LAGR_r,\; i\LAGR_i$ contribute to $T^\gamma_h$. Of course, then it is sufficient to consider only e.g. the variation of
$ \VAR e^h_\gamma $, because the second leads to the same eqs.

\section{Generalized \SC{Lagrangian} in tetrad-form}
\label{sec-gen-lagr}
To be able to make comparisons between all possible tetrad theories and to find a general expression of $T^\gamma_h$, 
we generalize the {\sc Lagrangian} of eq. \refeq{eq-lagr_z}
to a more general, real bilinear form of the $r^x_{ab}$, with constant factors $ H_{xy}^{abcd}$ 
\BE \label{eq-lagr-tetr-general}
   \LAGR = r^{x}_{ab} (r^y_{cd})^\ast\; H_{xy}^{abcd}.
\EE 
The expression $\LAGR$ is real ($\LAGR = \LAGR^\ast$) for arbitray complex $r^{x}_{ab}$, if and only if $H_{xy}^{abcd} = (H_{yx}^{cdab})^\ast$ holds.\\
We discuss here a general, \SC{Lorentz}-invariant, bilinear form\FN{
  This is not the most general form for complex $r^a_{bc}$. There exists e.g. a second, parity violating term
  ${\LAGR_{e} =  r^{x}_{ab} (r^y_{cd})^\ast \eta_{xy} \Delta^{abcd}}$, which is not used here.
} that contains four free, constant parameters $a,b,c,d$\FN{
  Please, do not mix indices and parameters. Variable indices can never occur as factors.
} 
and can be constructed with $\eta$, $\delta$ and $\Delta$
\BE \label{eq-def-H}
  H_{xy}^{abcd} = \eta^{ac}(a \delta_x^b \delta_y^d + b\delta_x^d \delta_y^b  + c\eta_{xy} \eta^{bd}) +
  d\; i(\eta_{fy}\delta^b_x\Delta^{afcd} - \eta_{fx}\delta^d_y\Delta^{cfab}) 
\EE
The last condition forces, that all four parameters $a,b,c,d$ must be real.
Every specific set of parameters $(a,b,c,d)$ describes a different theory. Because the Lagrangian is a simple sum, it 
can also be written as\FN{
  The summands $\LAGR_a,\dots,\LAGR_d $ are defined by eqs. \refeq{eq-lagr-tetr-general} and \refeq{eq-def-H},
  and we note the correspondence $\LAGR_d \equiv -2i\LAGR_i$ from comparing it with the definition of 
  $\LAGR_i$ in eq. \refeq{eq-lagr_z}.
}
\BE \label{eq-Lag-abcde}
\LAGR(a,b,c,d) = a \LAGR_a +  b \LAGR_b +  c \LAGR_c +  d \LAGR_d.
\EE 
All individual terms $\LAGR_a,\dots,\LAGR_d $ are real for arbitrary complex $r^x_{ab}$. 
Some of these \SC{Lagrangian} terms are listed explicitly in the appendix \ref{ssec-append-lagr}.
Comparing \refeq{eq-Lag-abcde} with eq. \refeq{eq-lagr_z}, we see that the matrix Lagrangian is represented as 
$\LAGR_z = \LAGR(1, -1, 0, -\frac 12)$.
In section \ref{sec-ricci-tetr} is shown, that also the \SC{Lagrangian} of the \SC{Einstein}-theory  $\LAGR_E \simeq R$
can be expressed by this formula as $\LAGR_E = \LAGR(1,- \frac 12,-\frac 14,0)$.

Similar decompositions of the \SC{Lagrangian} into a sum of terms,
mostly in the teleparallel context, can be found in \cite{SHIRAFUJI} and \cite{MEI} (eq. (17) there).
Also Itin \cite{ITIN-2}, following the coframe description, gives a 3-term decomposion (eqs. 3.3 - 3.16) as {\em most general form}, 
which is for real $r^a_{bc}$ equivalent to the first three terms in our eqs. \refeq{eq-def-H} - \refeq{eq-var-lagr-r}.
Of course, none of them has a parity violating (PV) term $\sim \LAGR_d$, because for real tetrads obviously holds $\LAGR_d = 0$.\\
However, a similar term  $\LAGR_{PV} =  r_a t^a$ with real tetrads was discussed in \cite{MUELLER-1983} as a 
possible cure for the initial value problem mentioned in sec \ref{sec-str-tetr}. But later it was shown that this term has to be rejected, because
it leads to a ghost for the linearized theory (\cite{KUHF}, p. 1219 and \cite{MUELLER-1985}, p. 751). For the complex theory, presented here,
the situation is quite different, because of the factor $i$ the terms decouple (for all test cases with real tetrads), 
as demonstrated in sections \ref{ssec-ssol} and \ref{ssec-ppn}. A deeper analysis of this, in connection with the discussion 
of the possibility of real tetrads, should be left to future work.\\
 
Now we derive the geometric stress-energy tensor from the general form in eq. \refeq{eq-lagr-tetr-general} by 
variation of the  tetrads $\VAR e^a_\mu$. As usual, we consider $ e^a_\mu $ and  $ (e^a_\mu)^\ast $ as independent functions.
Consequently we only have to variate  $r^{x}_{ab} $ and then express $\VAR r^x_{ab}$ in terms of $\VAR e^a_\mu$.\FN{
  The  variation of $\VAR e^a_\mu$ does not affect  
  $(r^x_{ab})^\ast$ because it is constructed  $ (e^a_\mu)^\ast $ and their inverses only.
  On the other hand, the variation of the complex conjugated $ \VAR (e^a_\mu)^\ast $ gives the same eqs.
}
So one gets as variation simply
\BE \label{eq-var-lagr-r}  \label{eq-def-ufba}
   \VAR \LAGR =  \VAR r^x_{ab}\; \underbrace {(r^y_{cd})^\ast H^{abcd}_{xy}}_{\DEF  U^{ab}_x } =   
   \VAR r^x_{ab} U^{ab}_x = \frac 12 \VAR r^x_{ab} U^{[ab]}_x.
\EE
We have defined here a new fundamental symbol $ U^{ab}_x \DEF (r^y_{cd})^\ast H^{abcd}_{xy} $. It is a linear
form of the $(r^y_{cd})^\ast$ with constant coefficients. Because $r^x_{ab}$ is antisymmetric with respect to the lower indices,
only the antisymmetric part  $ U^{[ab]}_x \DEF U^{ab}_x - U^{ba}_x $, which has also 24 components, 
is needed in \refeq{eq-var-lagr-r}. In the following is shown, how the 
stress-energy tensor is to compute using this symbol. Its explicit form, i.e. the
form of the factors $ H^{abcd}_{xy}$ and the constants $(a,b,c,d)$ are not needed for those general derivations.

\subsection {Stress-energy tensor for the generalized \SC{Lagrangian}}
For the computation of $T$ from the \SC{Lagrangian} $\LAGR$, 
the variation of all terms of the action integral must be expressed by the variations of covariant tetrads $\VAR e^a_\mu$, 
so we need the formulas for the inverse (contravariant) tetrads and the absolute value of the tetrad determinant 
$||e|| \DEF \sqrt{|e||e|^\ast} = \sqrt{-g}$, which are derived
from the orthogonality relations:
\BE \VAR e^\beta_b = - e^\beta_a e^\alpha_b \VAR e^a_\alpha\MBOX{and} \VAR |e| = |e| e^\gamma_h \VAR e^h_\gamma.
\EE
Inserting this, one gets 
\BEN  \label{eq-var-rafb}
  \VAR r^a_{fb} &= & \VAR(e^a_{[\mu,\alpha]} e^\mu_f e^\alpha_b) \nonumber
  = e^a_{[\mu,\alpha]} ( \VAR e^\mu_f e^\alpha_b +  e^\mu_f\VAR e^\alpha_b) + \VAR e^a_{[\mu,\alpha]} e^\mu_f e^\alpha_b\\
  &=& - e^a_{[\mu,\alpha]}(e^\mu_h e^\gamma_f e^\alpha_b +  e^\mu_fe^\alpha_h e^\gamma_b)\VAR e^h_\gamma
  + \VAR e^a_{\mu,\alpha} (e^\mu_f e^\alpha_b -e^\mu_b e^\alpha_f) ) \nonumber \\
  &=& \VAR e^h_\gamma (r^a_{hf} e^\gamma_b - r^a_{hb} e^\gamma_f)  
  + \VAR e^a_{\mu,\alpha} (e^\mu_f e^\alpha_b -e^\mu_b e^\alpha_f) 
\EEN
and the total variation of the action integral becomes
\BE \label{eq-var-int-lagr-tetr}
 \VAR I = \int d^4x \VAR (||e||\LAGR) = \int d^4 x (\VAR ||e||\LAGR + ||e|| \VAR \LAGR) =
   \int d^4x ||e||(\VAR e^h_\gamma (\frac 12 e^\gamma_h \LAGR + A^\gamma_h ) + \VAR e^h_{\gamma,\alpha} B^{\gamma\alpha}_h).
\EE
The here introduced new expressions 
$ A^\gamma_h \DEF \PD[\LAGR] {e^h_\gamma}$ and $B^{\gamma\alpha}_h \DEF \PD[\LAGR] {e^h_{\gamma,\alpha}}$ 
are to compute by inserting the eq. \refeq{eq-var-rafb}  into eq. \refeq{eq-var-lagr-r}, which expresses them by $U^{[fb]}_a$ :
\BE \label{eq-def-a-gh}
   A^\gamma_h =  (r^a_{hf} e^\gamma_b - r^a_{hb} e^\gamma_f) U^{fb}_a = 
     r^a_{hf} e^\gamma_b (U^{fb}_a - U^{bf}_a) =    r^a_{hf} e^\gamma_b U^{[fb]}_a
\EE
and the second is obviously the antisymmetric expression $B^{\gamma\alpha}_h = -B^{\alpha\gamma}_h$:
\BE \label{eq-def-b-gah}
  B^{\gamma\alpha}_a = (e^\gamma_f e^\alpha_b -e^\gamma_b e^\alpha_f)  U^{fb}_a = e^\gamma_f e^\alpha_b U^{[fb]}_a .
\EE
As usual, the variation term $\VAR e^h_{\gamma,\alpha} $ in eq. \refeq{eq-var-int-lagr-tetr} is eliminated by
partial integration (and neglecting the remaining surface integral) and this leads to the definition
of the {\em gravitational stress-energy tensor}, here written as $T^\gamma_h$:
\BE
\VAR I = \int d^4x ||e|| \VAR e^h_\gamma 
\big[\frac 12 e^\gamma_h\LAGR + A^\gamma_h - \frac 1{||e||} (||e||B^{\gamma\alpha}_h)_{,\alpha} \big] 
\DEF \int d^4x ||e||\; \VAR e^h_\gamma\; T^\gamma_h ,
\EE
with
\BE \label{eq-def-tgh}
 \underline{  T^\gamma_h = \frac 12 e^\gamma_h\LAGR + A^\gamma_h - \frac 1{||e||} (||e|| B^{\gamma\alpha}_h)_{,\alpha}} \;\;. 
\EE
This form with mixed-type indices (spacetime/tetrad, upper/lower) naturally arises
from tetrad variation. If we want to transform it into a homogenous representation, we have to use a convention
about the order of indices. Here we define $T^{\mu\gamma} \DEF e^{\mu h} T^\gamma_h$, i.e. the tetrad index should
become the first.
  
The above derivation of $T$ from $\LAGR$ is similar to the \SC{Einstein}-theory (\SC{Hilbert} 1915), except that we used
a more general \SC{Lagrangian} and tetrads instead of the metric tensor. For the \SC{Einstein}-case 
with $\LAGR_E = R(g_{\mu\nu},g_{\mu\nu,\lambda})$ and $\VAR \LAGR = \VAR g_{\mu\nu} T^{\mu\nu}_{(E)}$ 
it is easy to show, that one would obtain by tetrad variation like above, the tensor
$T^\gamma_h = e_{h\nu} T^{\nu\gamma}_{(E)}$, which is equivalent.\\
 
In the next section it will be shown, that a conservation law can be derived for the general 
stress-energy tensor 
defined in eq. \refeq{eq-def-tgh}, that expresses energy-momentum conservation. 

However, for the general theory, in contrast to \SC{Einstein}-theory, where 
$T^{\gamma\lambda}_{(E)} = R^{\gamma\lambda} - \frac 12 g^{\gamma\lambda} R$ holds, 
the symmetry and reality of $T^{\gamma\lambda}$ is not
guaranteed in all cases. This will be discussed in section \ref{sec-compare}.
\subsection{Energy-momentum conservation}
\label{sec-enmom-conserv}
In this section we derive a conservation law for the stress-energy tensor defined in eq. \refeq{eq-def-tgh}.
As explained above, this definition holds for all gravitation theories, which are derived
from a \SC{Lagrangian} of the form \refeq{eq-lagr-tetr-general}, including \SC{Einstein}- and matrix-theory.

 The easiest way to compute the covariant derivative is to use a tensor density (see e.g. \cite{EINSTEIN}),
 which here is defined
by (the tetrad index "$h$" has to be transformed into a spacetime index ``$\sigma$'')
\BE 
  {\mathcal T}^\gamma_\sigma \DEF ||e|| e^h_\sigma T^\gamma_h = 
  ||e||\big(\frac 12 \delta^\gamma_\sigma\LAGR + e^h_\sigma A^\gamma_h\big) -  e^h_\sigma(||e|| B^{\gamma\alpha}_h)_{,\alpha}.
\EE
We have to compute the divergence of this tensor density:\FN{
  Consider the antisymmetry of $B^{\gamma\alpha}_h = -B^{\alpha\gamma}_h$ and the relation 
  $e^h_\sigma A^\gamma_h = - e^h_\sigma e^b_\alpha r^a_{bh} B^{\alpha\gamma}_a = 
  e^h_{[\sigma,\alpha]} B^{\alpha\gamma}_h$ 
  (derived from eqn. \refeq{eq-def-a-gh}, \refeq{eq-def-b-gah}).
  Also used is the derivation of the tetrad determinant: $|e|_{,\sigma} = |e| e^\gamma_h e^h_{\gamma,\sigma}$
  and in eq. \refeq{eq-econs-4} the definition of $T^\alpha_h$ is inserted again.
}
\BEN
   {\mathcal T}^\gamma_{\sigma,\gamma} &=& 
   \frac 12 (||e||\LAGR)_{,\sigma} + (||e||e^h_\sigma A^\gamma_h)_{,\gamma} 
   -  e^h_{\sigma,\gamma}(||e|| B^{\gamma\alpha}_h)_{,\alpha} -  
   e^h_\sigma\underbrace{(||e|| B^{\gamma\alpha}_h)_{,\alpha\gamma}}_{=0} \\
   &=& \frac 12 (||e||\LAGR)_{,\sigma} + (||e||e^h_{[\sigma,\alpha]} B^{\alpha\gamma}_h )_{,\gamma} 
   -  e^h_{\sigma,\gamma}(||e|| B^{\gamma\alpha}_h)_{,\alpha} \\
   &=& \frac 12  (||e||\LAGR)_{,\sigma} + 
   (||e||e^h_{[\sigma,\alpha]} B^{\alpha\gamma}_h - e^h_{\sigma,\alpha} ||e|| B^{\alpha\gamma}_h)_{,\gamma} 
   + \underbrace{e^h_{\sigma,\gamma\alpha}(||e|| B^{\gamma\alpha}_h) }_{ = 0} \\
   &=&  \frac 12 (||e||\LAGR)_{,\sigma} - (||e||e^h_{\alpha,\sigma} B^{\alpha\gamma}_h)_{,\gamma} = \label{eq-econs-4}
   \frac 12 (||e||\LAGR)_{,\sigma} - e^h_{\alpha,\sigma\gamma} ||e||B^{\alpha\gamma}_h
   - e^h_{\alpha,\sigma} \!\!\!\! \underbrace{(||e|| B^{\alpha\gamma}_h)_{,\gamma}}_{
     = ||e||(\frac 12 e^\alpha_h\LAGR + A^\alpha_h - T^\alpha_h)} \\
   &=& ||e||(\frac 12 \LAGR_{,\sigma} +  \frac 14 (e^\gamma_h e^h_{\gamma,\sigma} + (e^\gamma_h e^h_{\gamma,\sigma})^\ast)  \LAGR 
   - e^h_{\alpha,\sigma\gamma} B^{\alpha\gamma}_h
   - e^h_{\alpha,\sigma} (\frac 12 e^\alpha_h\LAGR + A^\alpha_h - T^\alpha_h))\\
   &=& ||e||(\frac 12 \LAGR_{,\sigma}  +  \frac 14 ((e^\gamma_h e^h_{\gamma,\sigma})^\ast -e^\gamma_h e^h_{\gamma,\sigma} )  \LAGR 
   - e^h_{\alpha,\sigma\gamma} B^{\alpha\gamma}_h - 
   e^h_{\alpha,\sigma} (A^\alpha_h - T^\alpha_h)) \; . \label{eq-econs-x}
\EEN
If we use the condition, that $\LAGR(e^h_\gamma, e^{h\ast}_\gamma, e^h_{\alpha,\gamma}, e^{h\ast}_{\alpha,\gamma})$ does not {\em explicitly} 
depend on 
$x^\mu$, we can compute its partial derivation with the definitions of the terms $A,B$ 
\BE 
  \LAGR_{,\sigma}  = 
  \PD[\LAGR]{e^h_\gamma} e^h_{\gamma,\sigma} +  \PD[\LAGR]{e^h_{\alpha,\gamma}} 
  (e^h_{\alpha,\gamma})_{,\sigma} + cc. 
  = A^\gamma_h e^h_{\gamma,\sigma} +(A^\gamma_h)^\ast (e^h_{\gamma,\sigma})^\ast + 
  B^{\alpha\gamma}_h e^h_{\alpha,\gamma\sigma} + (B^{\alpha\gamma}_h)^\ast (e^h_{\alpha,\gamma\sigma})^\ast.
\EE
Inserting this in eq.  \refeq{eq-econs-x}, we compute the real part of the expression, where only one term on the rhs. remains:\FN{
  consider $\LAGR = \LAGR^\ast $
}
\BE \label{eq-t-gsg}
  \underline{\Re ({\mathcal T}^\gamma_{\sigma,\gamma})} = ||e|| \Re( e^h_{\alpha,\sigma} T^\alpha_h )=  
  \underline{\Re( e^h_{\alpha,\sigma} e^\mu_h \mathcal T^\alpha_\mu) } \; .
\EE
At last, we can easily show from the definition of the \SC{Christoffel} symbols (considering only real tetrads), 
that for any symmetric tensor $T^{\lambda\alpha} =T^{\alpha\lambda} $ holds
\BE g_{\lambda\mu} \Gamma^\mu_{\alpha\sigma} T^{\lambda\alpha} =  e^h_{\alpha,\sigma} e_{h\lambda}  T^{\lambda\alpha}, 
\EE
so  finally, if $T$ is symmetric and real, the covariant derivative vanishes
\BE \underline{{\mathcal T}^\gamma_{\sigma;\gamma} = 
  {\mathcal T}^\gamma_{\sigma,\gamma} - \Gamma^\gamma_{\sigma\beta} {\mathcal T}^\beta_\gamma = 0 } \; .
\EE
As conclusion it is to state, that 
the divergence of the stress-energy tensor is {\em zero, if $T$ is symmetric and real}.
This holds for all theories described by the \SC{Lagrangian} of eq \refeq{eq-lagr-tetr-general}, since the explicit structure of the
symbol $U^{[fb]}_a$ is not used in the above computation.

We have to add, however, that for spaces which represent {\em real matter distributions}, the actual symmetry
follows from the fact, that it equals the stress-energy tensor of matter $T_{(g)}^{\mu\nu} = T_{(m)}^{\mu\nu}$.
This equation is usually derived by simply adding both \SC{Lagrangians} and it postulates that matter acts as the source
of the gravitational spacetime curvature. Esp. for the cosmological most relevant cases, the 
ideal fluid approximation for matter is used, which is given by the real, symmetrical tensor
\BE T_{(m)}^{\mu\nu} = (\rho + p) u^\mu u^\nu + p g^{\mu\nu},
\EE
where $\rho$ is mass-energy density, $p$ pressure and $u^\mu$ the 4-velocity. This is discussed in detail in section
\ref{sec-compare}.

\subsection{Curvature scalar $R$ and \SC{Einstein-Lagrangian} in tetrad-form}
\label{sec-ricci-tetr}

In this section we will show, that the \SC{Lagrangian} of the \SC{Einstein}-theory can be written 
as special case of $\LAGR(a,b,c,d)$ in eq. \refeq{eq-lagr-tetr-general}.
To prove this, we have to express the curvature scalar $R$ by tetrads 
(we consider only real tetrads here).\\
This is a quite lengthy computation, because one has to start with the complete \SC{Riemann}-tensor, expressed by tetrads, 
and then to reduce it with $R = \eta^{mn} R^s_{mns}$. Similar computations can be found, with different notations, in various
papers, e.g. \cite{MOLLER}.
Therefore, we have put it into the appendix \ref{ssec-append-1} and will give here
the result (again  $r_b \DEF r^a_{ba}$ as contracted form)
\BE
  R = -2 e^{\lambda b} r_{b,\lambda}  + 
  \frac 14 r^c_{ab} \eta^{sb} (2r^a_{cs} + \eta_{xc}\eta^{an} r^x_{ns}) + \eta^{xb}r_x r_b \; .
\EE
This expression contains also second derivatives of the tetrads, namely the first term $r_{b,\lambda}$.
For the \SC{Lagrangian} it is eliminated by partial integration:
\BE
  \int |e| e^{\lambda b} r_{b,\lambda} =  - \int (|e| e^{\lambda b})_{,\lambda} r_b
  = - \int \eta^{sb} (|e| e^{\lambda}_s)_{,\lambda}  r_b =  \int |e| \eta^{sb} r_s r_b \;, 
\EE
so we finally get an \SC{Einstein-Hilbert-Lagrangian}, which is bilinear in the first derivatives $r^a_{bc}$:
\BE  \label{eq-EH-Lagr-tetr}
  \LAGR_E =  \eta^{sb} r_s r_b -   \frac 14 r^c_{ab} \eta^{sb} (2r^a_{cs} + \eta_{xc}\eta^{an} r^x_{ns})
  =  r^a_{fb} r^c_{gd} \eta^{fg}  (\delta^b_a \delta^d_c - \frac 12\delta^b_c \delta^d_a - \frac 14\eta_{ac}\eta^{bd} ).
\EE
The same expression, in different notation, can be found in \cite{MALUF-2001}, eq. (1) and  \cite{MALUF}, eq. (6).
Comparing this with eqs. \refeq{eq-lagr-tetr-general} - \refeq{eq-Lag-abcde} gives 
\BE \underline{\LAGR_E = \LAGR(1,-\frac 12,-\frac 14, 0)}.
\EE

\subsection{Isotropic coordinates and ``viable'' tetrad theories}
\label{sec-iso-sm}
The term ``viable'' gravity theories is widely used in the literature.
 Nester \cite{NESTER} (introduction),
defines it as ``one-parameter class of teleparallel theories which agree with Einstein's theory to post-Newtonian order''.
Muench et. all \cite{HEHL-1998}, give a similar definition of viable Lagrangians (p. 15), based on a 
three-parameter-set $(a_1,a_2,a_3)$, which is obviously equivalent to our set $(a,b,c)$.\FN{
  They give as viable class $a_1 = 1, a_2 = - 2, a_3 = arbitrary$ ($a_3 = - \frac 12$, for the teleparallel 
  equivalent of Einstein's theory).
}

In this section we give a classification of tetrad theories defined by eq. \refeq{eq-Lag-abcde}.
We show, that for all spacetimes, where {\em isotropic coordinates} can be used, 
a certain subset, described by the relation $a+b+2c = 0$ (including \SC{Einstein}- and matrix-theory), 
have the same stress-energy tensor. 
Consequently, they have the same vacuum solutions, e.g. the fundamental \SC{Schwarzschild}-metric
for spherical symmetry. Only those are considered as ``viable'' theories in the following sections.
All others fail in the reality test.\\
We request, that all viable theories {\em must have real tetrads} as solutions representing the \SC{Schwarzschild}-metric.
Thus we can neglect the term $d \LAGR_d $, which is zero for real tetrads, in this section. 
(Its variation produces {\em additional} imaginary terms $\sim i$, however, which have to vanish independently, see section \ref{ssec-ssol}.)\\ 
For concrete computations with tetrads, one must be careful not to mix the different index-types.
Therefore we introduce here the symbol $z^\mu_a \equiv e^\mu_a$ as replacement term for the inverse tetrads, in this 
and the next sections.\\
 
A static, isotropic coordinate system is defined with two real functions $f(x_1,..,x_3),g(x_1,..,x_3)$ 
and the diagonal tetrads
\BE\label{eq-sm-tetrads}
  (e^a_\mu) = diag[f,g,g,g],\quad (z^\mu_a)  = diag[\frac 1f, \frac 1g,\frac 1g,\frac 1g],   \quad |e| = fg^3,
\EE 
and it leads to the diagonal metric $ (g_{\mu\nu}) = diag[ f^2, -g^2, -g^2, -g^2] $, which includes the \SC{Schwarzschild}-metric.
The not vanishing derivatives of the tetrads are
\[ e^0_{0,k} = f_{,k},\quad e^k_{k,m} = g_{,m}, \quad k,m = 1,2,3  
\] 
We now substitute $f = \exp(\mu) $ and $ g = \exp(\lambda)$.
The non-vanishing antisymmetric forms $r^a_{bc}$ are ($k$ = fixed, all ``r-triplets'' - with three different indices - are zero):
\[ r^0_{0k}  = e^0_{[0,k]} z^0_0 z^k_k = \frac  {\mu_{,k}}{g},     \quad  
  r^k_{km}  = e^k_{[k,m]} z^k_k z^m_m =  \frac  {\lambda_{,m}}{g},  \quad k\ne m
\]
To compute the stress-energy tensor, we need the terms $U^{fb}_a$ defined in eq. \refeq{eq-def-ufba}, which are
\[
  U^{fb}_a  = \eta^{fg}( a r^c_{gc}\delta^b_a + br^b_{ga} + cr^c_{gd}\eta_{ac}\eta^{bd} )
\] 
and the non-zero antisymmetric forms are ($m$ = fixed)
\BEN \label{eq-tetr-u-fba}
  U^{[0k]}_0 &=& a r_k - (2c+b) r^0_{0k}  = -\frac 1g [(a+b+2c)\mu_{,k} + 2a\lambda_{,k} ] \\  
 U^{[mk]}_m &=&  a r_k - (2c+b) r^m_{mk}  = -\frac 1g [a\mu_{,k} + (2a+b+2c) \lambda_{,k} ],\quad m \ne k  \; . \nonumber
\EEN
From the combination of the $abc$-factors above, one can see, that all real tetrad theories with $ a+b+2c = 0$
have the same $U$-terms (up to a constant factor $a$, which we can set to $a=1$, without loss of generality). 

Since the constants $(a,b,c)$ appear nowhere else in 
the \SC{Lagrangian}, those theories have the same stress-energy tensor. In section \ref{sec-ricci-tetr} it is shown, that the
\SC{Einstein}-\SC{Lagrangian} is $\LAGR_E = \LAGR(1,-\frac 12, -\frac 14,0)$, which fulfills this criterion.
Also matrix-theory $\LAGR_z = \LAGR(1,-1,0, -\frac 12)$ belongs to this class.
\label{page-no-doublq}
Because $a+b+2c$ is the weight of all ``r-doublet''-quadrats in eq. \refeq{eq-Lag-abcde}
(terminology introduced in section \ref{sec-gen-lagr} eq. \refeq{eq-def-rafb}), 
this class is characterized by
\SC{Lagrangian}s, which do not contain quadrats of r-doublets.\FN{
  We note, that the matrix-\SC{Lagrangian} can also be characterized as the only one, that contains no quadrats of ``r-triplets'' neither.
}
It is generated by setting $ b = -1-2c$, which defines the set of ``viable'' theories 
by two real constants $c,d$
\BE \label{eq-lagr_v}
   \LAGR_v(c,d) \DEF \LAGR(1,-1-2c,c,d) = \LAGR_a -\LAGR_b +c\underbrace{(\LAGR_c -2\LAGR_b)}_{\DEF \LAGR_x} + d\LAGR_d \; .
\EE
This class is investigated in the following section \ref{sec-compare}. The value of the parameter $c$ then defines the
theory: $c=0$ (and $d = -\frac 12$) describes matrix-theory and $c = -\frac 14$ (and $ d = 0$) is the \SC{Einstein} theory.\FN{
  \label{fn-omit-d}
  The parameter $d$ is not explicitly implemented, because it suffices to omit all terms $\sim i$ 
  (or formally set $i=0$) to use $d = 0$. For $d\ne 0$, its actual value plays no roll for real, symmetric matter tensors (e.g. vacuum) as
  it is demonstrated in the various test cases in section \ref{sec-compare}. However, the matrix Lagr. forces $d = -\frac 12$.
}
For a convenient checking of the results, all terms of \refeq{eq-lagr_v} and 
the $U$-terms for this \SC{Lagrangian} are explicitely listed in  the appendix \ref{ssec-append-lagr}. 

In generalization of the eqs. \refeq{eq-tetr-u-fba}, it can be shown (by the structure of $\LAGR_x$), 
that for tetrad fields, where all ``r-triplets'' ${r^x_{yz} = 0,}\; {(x\ne y\ne z)}$ are zero,
the $U$-terms of all viable theories are equal (independent of ``$c$'') and consequently the stress-energy tensor is equal
to the \SC{Einstein}-tensor (except terms from $\LAGR_d$, of course). 
 
As the computations in the sections \ref{ssec-lin-pn} - \ref{sec-ppn2} show, these viable theories also agree with
the Einstein-theory in first and second PPN-order.

\section{Comparison between \SC{Einstein}- and matrix-theory}
\label{sec-compare}
For the comparison of different theories we use the general ``viable'' \SC{Lagrangian} $\LAGR_v(c)$, defined in eq. \refeq{eq-lagr_v}
above. For this we derive from eqs. \refeq{eq-def-H} and \refeq{eq-def-ufba}
the following $U$-terms, which are explicitly listed in the appendix eq. \refeq{eq-U-Lv} for convenient checking
\BEN 
  U^{[ab]}_x &=& (\eta^{ac}\delta^b_x - \eta^{bc}\delta^a_x) r_c^* - (1+2c) (\eta^{ac} (r^{b}_{cx})^* - \eta^{bc} (r^{a}_{cx})^*) 
  + 2c \eta^{ac} \eta^{bd} \eta_{xy} (r^{y}_{cd})^*  \nonumber
  \label{eq-U-Lv-c}
  \\ && + i (\delta^a_x t^{b*} - \delta^b_x t^{a*}) + i \Delta^{cfab} \eta_{x f}  r^*_c \; .
\EEN
The computation of the stress-energy tensor for all test cases is then done with the following steps.\\
\label{pg-steps-A-G}
{\bf (A)} We start with the 16 covariant tetrads $e^a_\mu$ which represent the problem and compute {\bf (B)} the determinant $|e|$ and
{\bf (C)} the 16 inverse tetrads $e^\mu_a$, defined by the orthogonality eq. \refeq{eq-tetr-orthogon}. 
{\bf (D)} compute the 24 coefficients
$r^a_{bc} \DEF e^a_{[\beta, \gamma]} e^\beta_b e^\gamma_c $. 
{\bf (E) } compute the 24 $ U^{[ab]}_x$ with above eq. \refeq{eq-U-Lv-c}
resp.  \refeq{eq-U-Lv}. 
{\bf (F) } compute $\LAGR = \frac 12 r^x_{ab} U^{[ab]}_x $ and the 16 $A^\gamma_h$ of eq. \refeq{eq-def-a-gh} and the 24 
$B^{\gamma\alpha}_a$ of eq. \refeq{eq-def-b-gah}. 
{\bf (G)} Finally compute $T^\gamma_h$ with eq. \refeq{eq-def-tgh} and optionally 
$T^{\mu\gamma} = \eta^{mh} e^{\mu}_m T^\gamma_h$. These components of the stress-energy tensor then contain the parameter ``$c$''
and are valid for the class $\LAGR_v(c)$.

For comparing the theories, we then have to use 
$c = 0$ for matrix theory and $ c = -\frac 14$ (and {\em formally} set $i = 0$) for the \SC{Einstein}-theory.

The above described computations are straightforward, but quite lengthy and error-prone. 
Existing software packages are either not well designed for these problems, or not free.\\
\label{page-symbolic}That is why, we have developed ``Symbolic'' \cite{Symbolic},  
a small Java-program for such symbolic formula manipulations and the test
of given solutions. It is a script-driven formula interpreter, especially designed for tensor calculus in GR, and produces
TeX- and PDF-output files.
It can be found, together with various sample scripts (nearly all test cases of this paper in tetrad formulation,
as well as the corresponding problems for the Einstein theory and their results as PDF-files). 
Also available on this server is a web interface for testing it.

\subsection{``Unphysical'' tetrads}
\label{sec-str-tetr}
In the literature this kind of tetrads are discussed since the 1980-ies and by some authors they are considered as 
``death warrant'' for the teleparallel theory (tetrad gravity). 
The first author, who presented them was Kopczy\'{n}ski. He showed in \cite{KOPCZ}, that for some metrices the field equations 
are insufficient to determine the tetrads (resp. torsion tensor) completely. Then followed several papers, which tried to
circumvent the problem, but all of them suffering from other serious physical problems \cite{MUELLER-1985,KUHF,CHEN}.
 Esp. Nester \cite{NESTER} gives a very good overview about this dilemma.
The essential statement of his paper is, however, that those tetrads are non-geneneric and occur only for very special solutions.
Later work, using Dirac's constraint algorithm showed, that generic initial values have deterministic evolution while 
certain special initial configurations allow some undetermined evolution possibly only within a limited spatial region \cite{CHENG}.

Here we show, that typical ``strange'' tetrads are excluded in the matrix theory, due to the parity violating term $\LAGR_d$ in 
eq. \refeq{eq-Lag-abcde}. 
A deeper, general analysis has to be done yet. 
A prototype for this kind of tetrads (compare \cite{NESTER}, p. 1008) 
is given with one arbitrary function $\chi(x^0)$:
\BE e^a_\mu = \MATRIC{\cosh \chi & 0 & 0 & \sinh\chi \\ 0 & 1 & 0 & 0 \\ 0 & 0 & 1 & 0 \\ \sinh \chi & 0 & 0 & \cosh\chi }, 
\qquad |e| = 1
\EE
and it produces a flat \SC{Minkowski}-metric $g_{\mu\nu} = \eta_{\mu\nu}$, if $\chi$ is real.\FN{
  I.e. $\chi = \chi^*$.
  For the matrix theory, however, we generally  consider $\chi(x^0)$ as complex valued function, which gives
  $g_{00} = -g_{33} = \Re(e^{\chi - \chi^*})$, $g_{03} = 0$, from the definition \refeq{eq-def-metrik}.
} 
The problem, that arose
within previous tetrad gravity theories was, that the vacuum field equations $T^\gamma_h = 0$ are {\em identically} fulfilled for any 
function $\chi(x^0)$ and thus do  not pose any restriction on it. This fact obviously contradicts the assumption, 
that the tetrads (resp. torsion) possess a physical meaning, because they cannot be derived from some initial conditions.\\ 
In the following we show, that for the matrix theory - also for solutions with real tetrads - 
there are non-vanishing terms $T^\gamma_h \ne 0$, accruing from $\LAGR_i$,
and thus this problem here does not exist.

We have only non-vanishing 2 $r$-terms, namely\FN{
  For the inverse tetrads we use again the symbols $z^\mu_a \DEF e^\mu_a$ to distinguish them from the $e^a_\mu$. 
  They are given as $z^0_0 = z^3_3 = \cosh \chi,\quad z^0_3 = z^3_0 = -\sinh\chi$.
  From this we get e.g. $ r^0_{03} = -e^0_{3,0} (z^0_0 z^3_3 - z^3_0 z^0_3) =   \chi_{,0} \cosh \chi$.
}
\BE r^0_{03} = -\chi_{,0} \cosh \chi, \qquad r^3_{03} = - \chi_{,0} \sinh \chi.
\EE
For the \SC{Einstein}-theory it is obvious (the curvature tensor is zero), 
that the vacuum field equations are identically fulfilled. Since here all ``r-triplets'' are zero, 
it is consequently already clear from the considerations of section \ref{sec-iso-sm}, 
that for real tetrads only  the variation of $\LAGR_d$ can contribute to the field equations.

For explicit computing, we do not list the intermediate $U$-terms here (10 of them $\ne 0$). 
We only note that $\LAGR = 0$, all 16 $A^\gamma_h = 0$ and 
10 $B^{\gamma\alpha}_a$ are $\ne 0$. Finally, four $T^\gamma_h$ do not vanish, which are explicitly
\BEN \label{eq-t-str}
  T^1_1 &=& \quad T^2_2 = \frac 12 (\chi_{,00}^* +(\chi_{,0}^*)^2) (e^{\chi-\chi^*} -  e^{\chi^*-\chi}) 
  +\frac 12 \chi_{,0}^*\chi_{,0} (e^{\chi-\chi^*} + e^{\chi^*-\chi}) - (\chi_{,0}^*)^2 e^{\chi-\chi^*} 
  \MBOX{and}   \nonumber \\
  T^1_2 &=&  -T^2_1 =  \frac i2 (\chi_{,00}^* +(\chi_{,0}^*)^2) (e^{\chi-\chi^*} +  e^{\chi^*-\chi}) 
  +\frac i2 \chi_{,0}^*\chi_{,0} (e^{\chi-\chi^*} - e^{\chi^*-\chi}) - i(\chi_{,0}^*)^2 e^{\chi-\chi^*}.
\EEN
We recognize from eq. \refeq{eq-t-str}: 
1. The constant ``$c$'' does not appear in any component of $T$, and consequently the real part of the stress-energy tensor is independent 
of ``$c$'', i.e. equal for all viable theories.\\ 
2. For real tetrads ($\chi = \chi^* $) follows $T^1_1 = T^2_2 \equiv 0$ and the other two components $T^1_2 \sim i,\;\; T^2_1 \sim i $ 
are {\em only present} for matrix-theory and the vacuum eqs. $T^1_2 = T^2_1 = 0$ pose restrictions on $\chi(x^0)$ only here.\\
3. The {\em unique} solution of $T^1_1 = T^1_2 \stackrel != 0$ for general complex $\chi$ (within the matrix-theory) 
is easily derived as simple linear function 
\underline {$\chi(x^0) = k x^0 + c$} with two constants: $k = real$, but complex $c$. 
The free parameters $c,k$ are then clearly determined by boundary conditions.\\
As bottom line we resume again, that the problem solved in this section was not the {\em existence} of a solution, but the {\em exclusion}
of physically unreasonable solutions.  Of course, our computations here are no ultimate proof, that such solutions 
do not exist, but a strong argument.

\subsection{\SC{Schwarzschild}-solution}
\label{ssec-ssol}
In this section we show, that the important \SC{Schwarzschild}-metric is also a vacuum solution of the matrix
field equations. From the considerations in section \ref{sec-iso-sm} it is clear, that the real parts of the stress-energy
tensor are equal for all viable theories, i.e. also for matrix theory. It remains to clarify, however, that the additional
imaginary terms do not pose unsolvable constraints.\\
The tetrads to use are the same as in eq. \refeq{eq-sm-tetrads}. It shows, that it suffices to use only real tetrads for simplicity, 
i.e. real functions $f(r),g(r)$. The computations (again following all steps from (A) to (G) on page \pageref{pg-steps-A-G}, 
which we do not list here) 
finally gives the following components of $|e|T^\gamma_h$ (we list here 5 representatives,
the other 11 are similar)
\BEN
|e|T^0_0 &=& 2(g_{,11} + g_{,22} + g_{,33}) - (g_{,1}^2 + g_{,2}^2 + g_{,3}^2)/g 
\\ \nonumber |e|T^0_1 &=& 2i(f_{,2} g_{,3} -f_{,3} g_{,2})/f  
\\ \nonumber |e|T^1_0 &=& -4i(f_{,2} g_{,3} -f_{,3} g_{,2})/g  
\\ \nonumber  |e| T^{1}_{1} &=& 
 (  f g_{,1}^2 - f g_{,2}^2 - f g_{,3}^2)  / g^2+ (  2 f_{,1} g_{,1} + f g_{,22} + f g_{,33})  / g + f_{,22} + f_{,33}
\\ \nonumber |e| T^{1}_{2} &=& 
 - f_{,12}+ (  f_{,2} g_{,1} - f g_{,12} + f_{,1} g_{,2})  / g + 2 f g_{,1} g_{,2} / g^2
\EEN
Their inspection shows, that imaginary terms $\sim i$ only occur for $T^0_k$ and $T^k_0$. They are zero for all 
spherically symmetric functions  $f(r),g(r)$, which was required.
All other terms are real, and - since independent of ``$c$'' - equal for all theories. Hence it is
obvious, that the vacuum solution is the well-kown \SC{Schwarzschild}-field. For completeness, we sketch some basic steps here. 
By the substitution $g = e^\lambda$ we get for $T^0_0= 0$ the simple second order eq. 
\BE  2(\lambda'' + \frac 2r \lambda') +\lambda'^2 = 0.
\EE 
This is solved by $ \lambda' = - \frac 2{r(1+ 2r/M)}$ and leads to 
the well-known expression with an arbitrary constant $c_1$:
\BE g_{kk} = -g^2 = -exp(2\lambda) = -c_1 (1+ \frac M{2r})^4.
\EE
The other components give two similar equations and finally lead to
\BE g_{00} = f^2 = exp(2\mu) = c_0 \frac{(1- \frac M{2r})^2}{(1+ \frac M{2r})^2}.
\EE
which is the known metric for isotropic coordinates \cite{MTW}, \cite{STEPHANI}.
According to the metric definition \refeq{eq-def-metrik} the signature $[+,-,-,-]$ is a 
{\em forced result} of the matrix theory. This is in contrast to other tetrad- or the Einstein-theory, where the signature must
be postulated as additional assumption (e.g. as boundary condition for $r \to \infty$).  
Unlike other tetrad theories, the matrix theory also does not presuppose the 
\SC{Minkowski} metric, when the fundamental matrix \SC{Lagrangian} 
eq. \refeq{eq-def-lagr-matr} is considered.

\subsection{PPN-test}
\label{ssec-ppn}
In this section we perform a comparison between Einstein- and matrix-theory, based on the well-known
PPN-scheme. It is shown, that both theories give identical results up to standard parametrized post-Newton (PPN) 
approximation order \cite{MTW}, \cite{WILL}.\FN{
  We use the flat spacetime metric with the signature $\eta = [1,-1,-1,-1]$, which has
  the opposite sign as in most GR-textbooks, and also for the metric results the opposite sign.
  Our form naturally evolves from the matrix theory (see eq. \refeq{eq-matr-norm}). 
  It is also the form mostly used in relativistic quantum mechanics.
}  
\subsubsection{Linear PN approximation}
\label{ssec-lin-pn}
For solving the {\bf linear field equations} of the PPN scheme, we use a tetrad ansatz with $2\times 3$ non-diagonal 
- generally complex valued - terms
$v_k \pm h_k\; (k= 1,2,3)$ and 3 equal space-diagonal elements $e^k_j = g\delta^k_j$, with $f,g \sim 1 +\ORD 2$ and 
$h_k, v_k \sim \ORD 3$. It produces the metric tensor, which is used for the linear PPN approximation.
Latin letters $j,k,..= 1,2,3$ denote space indices. 
The symbols used here are in accordance to those for the \SC{Schwarzschild} metric in eq. 
\refeq{eq-sm-tetrads} and section \ref{ssec-ssol}, because this ansatz can be considered 
as its generalization.
\[ e^0_0 = f = 1+ \mu,\quad e^0_k = v_k + h_k,\quad  e^k_0 = v_k - h_k, \quad  e^k_k = g = 1+\lambda
\]
\BE \label{eq-tetr-ppn-lin} 
  (e)   = \MATRIC { f & v_1+h_1 & v_2 + h_2 & v_3 +h_3\\ v_1 - h_1 & g & 0 & 0\\ v_2 - h_2 & 0 & g & 0 \\  v_3 - h_3  & 0 & 0 & g }.
\EE
and gives the linearized metric
\BEA 
  g_{00} &=& \Re((e^{0}_0)^* e^0_0 - (e^{1}_0)^* e^1_0-\cdots) \approx +1 +2\Re (\mu) \quad\quad\; \DEF +1 + h_{00} = +1 + \ORD 2,\\
  g_{11}  &=& \Re((e^0_1)^* e^0_1 - (e^1_1)^* e^1_1-\cdots) \approx -1 - 2\Re(\lambda)\quad\quad\; \DEF -1 + h_{11}= -1 + \ORD 2 \;, \dots \\
  g_{01} &=& \Re((e^0_0)^* e^0_1 - (e^1_0)^* e^1_1-\cdots) \approx \quad \quad  \; 2 \Re(h_1)\qquad \DEF\quad\quad \;\; h_{01} 
  = \ORD 3 \;, \dots \\  g_{12} &\approx& 0,\quad \dots.
\EEA
In this approximation only the real parts of $h_k$ enter into the metric, and the $v_k$ do not contribute at all.
For the computation of the stress-energy tensor in linear approximation, 
as inverse tetrads $e^\mu_a$ are to use the simple diagonals $e^\mu_a \approx \delta^\mu_a$ and for the determinant
$|e| \approx fg^3 \approx 1$. It shows in the following, that it suffices again to consider only  {\em real} tetrads, 
i.e. all $\lambda,\mu, h_k,v_k = real$. This ansatz is valid up to the requested approximation order and
solves all complex equations.

We receive the following $r$-terms as step (D) in the general scheme (six representatives listed here):
\BEN
&& r^0_{01} = - (h_1+v_1)_{,0} + \mu_{,1} \qquad
r^0_{12} = h_{1,2} - h_{2,1} +  v_{1,2} - v_{2,1}\\&&
r^1_{01} = (v_1-h_1)_{,1} - \lambda_{,0} \qquad\quad
r^1_{02} = (v_1-h_1)_{,2} \qquad \qquad r^1_{12} = \lambda_{,2} \qquad r^1_{23} = 0 \nonumber
\EEN
In the linear case, there is no need for the auxiliary terms $\LAGR, A^\gamma_h, B^{\gamma\alpha}_h$, since
the stress-energy tensor can be computed from the $U^{[fb]}_a$ directly as 
$T^\gamma_h \approx - B^{\gamma\alpha}_{h,\alpha} \approx - U^{[\gamma\alpha]}_{h,\alpha}$. 
Here we need the components $T^{\mu\gamma} \approx  \eta^{\mu h} T^\gamma_h $ which have to be equal to the ideal fluid
tensor of the matter $ T^{\mu\gamma}_M$.
For this comparison we compute the symmetrized  $T^{(\mu\gamma)} = T^{\mu\gamma} + T^{\gamma\mu}$ 
and antisymmetrized components $T^{[\mu\gamma]} = T^{\mu\gamma} - T^{\gamma\mu}$, 
for which we list six representatives here
\begin{eqnarray} 
\frac 12 T^{(0 0)} &=& 2 (\lambda_{,11} +  \lambda_{,22}+  \lambda_{,33}) + \ORD 4  \label{eq-T-00s}
\\\frac 12 T^{(1 0)} &=& \label{eq-T-10s}
 h_{1,22} + h_{1,33} - h_{2,12} - h_{3,13} -2 \lambda_{,01}  + i (v_{3,2}-  v_{2,3})_{,0}  + \ORD 5 
\\\frac 12 T^{(1 1)} &=& 
 2 h_{2,02} + 2 h_{3,03} + 2 \lambda_{,00}   - \lambda_{,22} - \lambda_{,33} - \mu_{,22}  - \mu_{,33}
+ 2 i (v_{2,3} - v_{3,2})_{,1}  +\ORD 5
\\\frac 12  T^{(2 1)} &=&  \label{eq-T-21s}
 - h_{1,02} - h_{2,01} + \lambda_{,12} + \mu_{,12}  + i (v_{3,1} -  v_{1,3})_{,1} +  i(v_{2,3} -  v_{3,2})_{,2}  +\ORD 5
\\\frac 12  T^{[1 0]} &=& \label{eq-T-10a}
 (  4 c + 1)  (v_{2,12}+  v_{3,13} -  v_{1,22} - v_{1,33})+ i (h_{3,2} - h_{2,3})_{,0}   + 2 i( v_{3,2} - v_{2,3})_{,0} +\ORD 5
\\\frac 12  T^{[2 1]} &=&  \label{eq-T-21a}
 (  4 c + 1)  (v_{2,01} -  v_{1,02})+ i(  h_{1,13} + h_{2,23} + h_{3,00} + h_{3,33})  + i (\lambda_{,03}   -  \mu_{,03})
 \\ && + i (v_{3,00} +  v_{3,11} +  v_{3,22} -  v_{3,33}   -2  v_{1,13} -2  v_{2,23}) + \ORD 5  \nonumber
 \end{eqnarray} 
The inspection of these terms shows the following.\\
1. The {\bf Einstein theory} is given by $c= -\frac 14 $ and formally $i=0$, which results in - as it must be - all $T^{[\mu\nu]} \equiv 0$.
The symmetric terms are simplified with the usual 4 gauge conditions, which read here
\[ \lambda_{,k}+ \mu_{,k} = \ORD 4 \MBOX{and} 2h_{k,k} + 3\lambda_{,0} = \ORD 5
\] 
The fact, that the field variables $v_k$ do not enter the field eqs., is a consequence of not contributing to the metric  in this
approximation. This is a basic problem of the - a priory symmetric - TEGR theory: 
the number of field variables exceeds the number of field eqs., resulting in free fields \cite{MOLLER}.

Since it is known, that the Einstein-theory gives the correct results, there is no need to compute it here.
We sketch here only the basic computation steps, needed in the following, after \cite{MTW,WILL}.
The matter tensor of a perfect fluid is in this approximation given as (with rest-mass density $\rho$, pressure $p$ and velocity $u_j$)\FN{ 
  The geometric stress-energy tensor $T^{\alpha\beta}$ in our notation equals $-8 \pi \times$ the matter tensor $T_M^{\alpha\beta}$, 
  and $u_j = u^j$ 
}
\BE
  T^{00} \stackrel != -8 \pi \rho, \qquad
  T^{j0}  \stackrel != -8 \pi\rho u_j, \qquad
  T^{jk}  \stackrel ! =   -8 \pi(\rho u_j u_k + p \delta^{jk})
\EE
The known result is 
\BE 
  \mu = \frac 12 h_{00} = - U + \ORD 4,\quad h_j = \frac 12 h_{0j} =  \frac 74 V_j + \frac 14 W_j + \ORD 5, 
  \quad
  \lambda = -\frac 12 h_{11} = U + \ORD 4, \dots
\EE
with the auxiliary fields:
\BE
 U \DEF \int \frac {\rho(t, \MV x') }{|\MV x - \MV x'|} d^3x',
 \quad V_j \DEF \int \frac {\rho(t, \MV x') u'_j }{|\MV x - \MV x'|} d^3x',
 \quad W_j \DEF \int \frac {\rho(t, \MV x') (\MV u' \cdot (\MV x - \MV x')) (x_j - x_j') }{|\MV x - \MV x'|^3} d^3x'.
\EE

2. For the {\bf matrix theory} we have to show, that all additional terms in eqs. \refeq{eq-T-10s} ... \refeq{eq-T-21a}
(compared to the Einstein-case above) are zero in the requested order, and consequently the metric is the same. 
The above computation has made no use of the fields $v_k$, which thus can be freely chosen 
(within the order limit $ v_k \sim \ORD 3$). We set it here as with a simple ansatz from one potential ``$v$''
\BE v_k = v_{,k}.
\EE
With this ansatz all additional summands in the symmetric eqs. \refeq{eq-T-10s} ... \refeq{eq-T-21s} vanish.
Also the term  $i (h_{3,2} - h_{2,3})_{,0} \sim \ORD 4$ in  \refeq{eq-T-10a} does not contribute in the required order $T^{10} \sim \ORD 3$
and finally the last eq. \refeq{eq-T-21a}\FN{
   Consider $ h_{3,00}, v_{3,00} \sim \ORD 5$ 
} 
\BE
\frac 12  T^{[2 1]} = i (h_{3,00} +  v_{3,00} + \frac 12\lambda_{,03} - \Delta v_{,3})  
\approx i(\frac 12\lambda_{,03} - \Delta v_{,3}) \stackrel != \ORD 5
\EE
leads to an additional ``gauge'' condition for the potential $v$
\BE \Delta v = \frac 12 \lambda_{,0}.
\EE
It is solved with the help of the ``super-potential'' $\chi(\MV x,t) \DEF - \int {\rho(t, \MV x') }{|\MV x - \MV x'|} d^3x'$ 
(defined in \cite{WILL}, p. 94)
and gives $v = - \frac 14 \chi_{,0}$ and finally
\[ v_j =  - \frac  14 \chi_{,0j} = -\frac 14 (V_j - W_j).
\]

As bottom line of this section we can state, that all $T^{\mu\nu}$ and consequently also the metric 
for the matrix theory are identical to those of the Einstein theory.
\subsubsection{Second PPN-order}
\label{sec-ppn2}
To perform the second order calculations we have to determine $g_{00} $ up to $\ORD 4$. 
For this we have to use the same ansatz \refeq{eq-tetr-ppn-lin} for $(e)$ with more accurate inverse and
the gauge conditions of section \ref{ssec-lin-pn} already implemented
\BE \label{eq-tetr-ppn-2}
  (e) = \left( \begin{array}{ccccc}
    f &  h_1 + v_{,1} &  h_2 + v_{,2} &  h_3 + v_{,3}\\ v_{,1}- h_1  &  1 / f & 0 & 0\\  v_{,2}- h_2  & 0 &  1 / f & 0\\ 
    v_{,3} - h_3  & 0 & 0 &  1 / f\\\end{array} \right)
    \quad 
    (z) = \left( \begin{array}{ccccc}
      1 / f &  - h_1 - v_{,1} &  - h_2 - v_{,2} &  - h_3 - v_{,3}\\ h_1 - v_{,1} &  f & 0 & 0\\ h_2 - v_{,2} & 0 &  f & 0\\ 
      h_3 - v_{,3} & 0 & 0 &  f\\\end{array} \right)
\EE
with $f = e^\mu$. The inverse tetrads $ (z) = (e)^{-1}$ above are sufficiently accurate up to  $\ORD 5$ and also $|e| = 1/f^2 + \ORD 6$ .
Again, all tetrads can be considered as real, also for this approximation order.
We skip all intermediate steps (D)...(G) here and give only the requested $T^{00} = \eta^{mh} z^0_m T^0_h $.

We obtain accurately up to order $\ORD 5$:\FN{ 
  The terms $2 i\mu_{,3} (h_{1,2}-h_{2,1}) +\cdots \sim \ORD 5 $ can also be neglected, because $\mu$ is needed only up to $\ORD 4$.  
}
\BEN  
   T^{00} &=&    \mu_{,1}^2 + \mu_{,2}^2 +  \mu_{,3}^2 - 2 (\mu_{,11} + \mu_{,22} + \mu_{,33}) 
   \\&& +   2i\mu_{,3} (h_{1,2}-h_{2,1})  +   2i\mu_{,2} (h_{3,1}-h_{1,3})  +   2i\mu_{,1} (h_{2,3}-h_{3,2})
   +  \ORD 6 \nonumber 
   \\ &= &    \mu_{,1}^2 + \mu_{,2}^2 +  \mu_{,3}^2 - 2 (\mu_{,11} + \mu_{,22} + \mu_{,33}) +\ORD 5.
\EEN
Again $T^{00}$ does not depend on the value of the parameter ``$c$'', i.e. it is identical for all theories of this class.
 It is consequently clear without computation, that matrix theory is up to this order
not distinguishable from the \SC{Einstein} theory.
\subsection{New vacuum solutions}
\label{new-vac-sol}
The aim of this section is, to present some new vacuum solutions, which  the  \SC{Einstein}-theory 
does not possess. In the light of section \ref{sec-str-tetr}, where we showed, that the matrix theory includes
 additional {\em constraints}, these extra degrees of freedom are not quite obvious. It might be possible, that
solutions of this type can help to solve the galaxy rotation problem without the 
obscure ``dark matter'', \cite{BROWST-MOFFAT,WIKI-GAL-ROT}.\\

We use the following simple, static tetrads, with all diagonal elements  $= 1$ and three real 
functions $v_k(x^1,x^2,x^3), k = 1,2,3 $. 
Here {\em exact} vacuum solutions can be computed quite easily, because $|e| = 1$ holds, 
and the inverse tetrads are also simple:
\BE \label{eq-tetr-new-sol} 
  (e)   = (e^a_\mu) 
  = \MATRIC { 1 & v_1 & v_2 & v_3\\ 0 & 1 & 0 & 0\\  0 & 0 & 1 & 0 \\  0  & 0 & 0 & 1 },\quad
  (e)^{-1}  = (z^\mu_a) 
  = \MATRIC { 1 & -v_1 & -v_2 & -v_3\\ 0 & 1 & 0 & 0\\  0 & 0 & 1 & 0 \\  0  & 0 & 0 & 1 }.
\EE
The resulting metric is
\BE g_{00} = 1, \quad g_{0k} = v_k, \quad g_{ik} = v_i v_k - \delta_{ik}. 
\EE
The only three non-vanishing $r^a_{bc}$ terms are from the definition eq. \refeq{eq-def-rafb}
\[ r^0_{12} = v_{1,2} - v_{2,1}, \quad r^0_{13} = v_{1,3} - v_{3,1}, \quad
r^0_{23} = v_{2,3} - v_{3,2}
\]
and the stress-energy tensor  $T^{\gamma}_h $ can be computed {\em exactly}, following the remaining steps (E),.., (G), 
with 5 repesentatives listed here (the 11 others similar):
\begin{eqnarray*} 
   T^{0}_{0} &=& 
   3 c (v_{1,2} - v_{2,1})^2 + 3 c (v_{1,3} - v_{3,1})^2 + 3 c (v_{2,3} - v_{3,2})^2 
   \\ &&   +  2 c (v_{1,22} +  v_{1,33}  -  v_{2,12}  - v_{3,13})  v_{1} 
   + 2 c (v_{2,11} +  v_{2,33} - v_{1,12}  - v_{3,23})  v_{2}
   \\ &&    + 2 c ( v_{3,11} +  v_{3,22} -v_{1,13} - v_{2,23})  v_{3}
   \\T^{0}_{1} &=& (  1 + 2 c) ( v_{2,12} + v_{3,13} -  v_{1,22} -  v_{1,33}) + 
   \\ &&
   c [  (v_{1,2} -v_{2,1})^2  +  (v_{1,3}- v_{3,1})^2 -   (v_{2,3} - v_{3,2})^2 ]  v_{1}  
   \\ && +2 c ( v_{1,3} -  v_{3,1})(v_{2,3}  -  v_{3,2})  v_{2}  
    + 2 c (v_{1,2}- v_{2,1})( v_{3,2} - v_{2,3})  v_{3}
   \\ &&  + i v_{1} (v_{2,3} -  v_{3,2})_{,1}  + i v_{2} (v_{3,1} - v_{1,3} )_{,1}
    + i v_3 (v_{1,2} - v_{2,1})_{,1}
   \\ T^{1}_{0} &=&  2 c (-v_{1,22} - v_{1,33} +  v_{2,12} + v_{3,13})
   \\ T^{1}_{1} &=& 
   - c (v_{1,2}-  v_{2,1})^2 - c (v_{1,3} -  v_{3,1})^2 + c (v_{2,3} - v_{3,2})^2   + i (v_{3,2} -  v_{2,3})_{,1}
   \\ T^{1}_{2} &=&    2 c (v_{2,3} - v_{3,2})(v_{3,1} - v_{1,3})+ i (v_{3,2} - v_{2,3})_{,2} 
\end{eqnarray*} 
We discuss here the vacuum solutions $T^\gamma_h = 0$. 
For the case $c\ne 0$ they force immediately
\BE  \label{eq-flat-v} 
 v_{1,2} -v_{2,1} =  v_{1,3} - v_{3,1} = v_{2,3} - v_{3,2} = 0
\EE
This is exactly the condition for the flat (Minkowskian) spacetime $r^a_{bc} = 0$, 
which is consequently {\em the only
vacuum solution} for the \SC{Einstein}-theory with $c = -1/4$.

For the matrix-theory we have instead $c=0$, where the situation is quite different.
Then the solutions of  $T^\gamma_h = 0$ are given by 
\BE \label{eq-mag-grav} (v_{k,j} - v_{j,k})_{,m} = 0, \MBOX{i.e.} v_{k,j} - v_{j,k} = c_{kj} = const., \qquad (k,j,m = 1,2,3),
\EE
which is obviously a generalization of eq. \refeq{eq-flat-v}.    
The three antisymmetric constants $ c_{km} = - c_{mk}$, which define an axial vector, 
offer new degrees of freedom as linear functions $v_k = \frac 12 c_{km}x^m $, that in the \SC{Einstein}-theory all must be zero. 
We have to discuss however, that these solutions - although exact -
cannot be regarded as global solutions, because the associated metric is not asymptotically flat.
But they could probably be considered as regional approximation of similar generalized tetrads, which have to be 
found yet.\\
In the realm, where the above solutions are approximately valid, they significantly modify e.g. the motion of test particles.
This should be shortly sketched by a simple example.
The relativistic equations of motion for small velocities $u^0 \approx 1, u^k \ll 1$ give approximately\FN{
  The relevant \SC{Christoffel}-symbols are  $\Gamma^m_{0k} \approx \frac 12 (g_{0m,k} - g_{0k,m}) = \frac 12 (v_{m,k} - v_{k,m}) $
}
\BE \dot u^m \approx (v_{m,k} - v_{k,m}) u^k
\EE
If the motion then is considered as rotation inside a plane, perpendicular to the axis $c_{km}$, 
we find a constant angular velocity, i.e. the tangential velocity is proportional to the distance from the axis.
If we consider this as galaxy rotation, this increase is too fast, compared with the known flat 
rotation curves, \cite{BROWST-MOFFAT,WIKI-GAL-ROT}, 
but this could be surely attributed to the simplicity of the tetrad ansatz eq. \refeq{eq-tetr-new-sol}.     

\section{$U(1)$ Noether-current} 
Noether's theorem tells us, that every symmetry of the \SC{Lagrangian} leads to a conserved current. 
The simplest case for matrix theory is the abelian $U(1)$ symmetry $e^a_\mu \to e^{i\VP} e^a_\mu$, as explained on page
\pageref{Lagr-Symm}. This current has no counterpart in real tetrad theories. 
In \SC{Dirac}'s theory, however, it results in the conservation of charge \cite{EBERT}.\\
To derive it here, we define the complex tensor $I^\alpha \DEF ||e|| B^{\gamma\alpha}_h e^h_\gamma =  ||e|| e^\alpha_b U^{[fb]}_f$
(definition eq. \refeq{eq-def-b-gah}).
Inserting the field eqs. \refeq{eq-def-tgh} gives (with $T \DEF T^\gamma_h e^h_\gamma $):\FN{ 
  consider $ e_h^\gamma  e^h_\gamma = 4$ and $ A^\gamma_h e^h_\gamma = -2\LAGR$
}
\BEA
  I^\alpha_{,\alpha}  &=&  (||e|| B^{\gamma\alpha}_h e^h_\gamma )_{,\alpha}  =   
  (||e|| B^{\gamma\alpha}_h)_{,\alpha} e^h_\gamma  + ||e|| B^{\gamma\alpha}_h e^h_{\gamma,\alpha} 
  \\&=& ||e||( \frac 12 e^\gamma_h\LAGR + A^\gamma_h - T^\gamma_h) e^h_\gamma  + ||e|| B^{\gamma\alpha}_h e^h_{\gamma,\alpha} 
  =  ||e||( - T  + \frac 12 B^{\gamma\alpha}_h e^h_{[\gamma,\alpha]}) 
  \\&=& ||e||( - T  + \LAGR).
\EEA
Since $\LAGR = real$ (and also $T= real$ assumed) follows $\Im (I^\alpha)_{,\alpha} = 0$, and consequently  
is the {\em conserved real} $U(1)$-Noether-current to define as the imaginary part
$ J^\alpha \DEF \Im (I^\alpha) $.
By inserting the $U$-terms eq. \refeq{eq-U-Lv-c} for the matrix theory (with $c=0$, see also \refeq{eq-lagr_z} and \refeq{eq-U-Lv})
we get
\[ I^\alpha =  ||e|| e^\alpha_b U^{[fb]}_f = ||e||  e^\alpha_b (- 2 \eta^{bc} r_c^*  + 3  i t^{b *} ) 
\]
and finally the conserved real current
\BE \label{eq-J-alpha}
  J^\alpha =  \Im (I^\alpha) = \frac i2 ( I^{\alpha *} - I^\alpha ) 
  = ||e|| \big[ i \eta^{bc}  (e^{\alpha }_b r_c^* -e^{\alpha *}_b r_c)   
    + \frac 3 2 (t^{b*} e^\alpha_b   + e^{\alpha *}_b t^b)   \big].
\EE
Therein the first term vanishes for real tetrads (it contains only r-doublets) and the second term contains only r-triplets.\\
The physical interpretation of this current is yet unclear. Its explicit computation shows, that it is zero for
all astrophysical test cases in section \ref{sec-compare}, including PPN-tests up to order $ \leq \ORD 4$. 
Hence it can clearly not be identified with a macroscopic matter flow. 
However, $J^\alpha $ is not zero for the new vacuum solutions in section \ref{new-vac-sol}.

\section{When are real tetrads possible?}
\label{sec-RMT}
For comparison with existing ``real tetrad theories'', we discuss here a modified matrix theory, 
which is described by the \SC{Lagrangian} $\LAGR(1,-1,0,0) \equiv \LAGR_a - \LAGR_b$ in eq. \refeq{eq-Lag-abcde} 
(i.e. without the PV-term $\LAGR_d $, resp. $ d = 0$), 
and only considering {\em real tetrads}, for briefness labelled here as ``real matrix theory'' (RMT). 
Its $U$-terms are represented by setting $c = 0$, and formally  $i=0$ in  eq. \refeq{eq-U-Lv-c}. 
As explained in sec. \ref{sec-iso-sm}, 
it belongs to the set of viable theories, which is widely discussed in the literature \cite{HEHL-1998,NESTER}.
The gravitational field eqs. of this ``RMT'' (e.g. for the vacuum $T^\gamma_h = 0$) are then a set of 16 real eqs. 
for the 16 real tetrad components $e^a_\mu$.\\ 
The ``complex matrix theory'' - presented in this paper - differs from this ``RMT'' by additional
terms $ T^\gamma_{(i)h} \sim i$ in the stress-energy tensor\FN{
  For general complex tetrads these terms $ T^\gamma_{(i)h} \sim i$ are not purely imaginary, 
  nor are the $T^\gamma_{(r)h}$ real. This is only the case for real tetrads. 
} in eq. \refeq{eq-def-tgh}, which originate from the variation of $\LAGR_d$ (as an example 
see the linear PPN-tensor in the eq-system \refeq{eq-T-00s} ... \refeq{eq-T-21a}).\\
If therein the tetrads are still constrained to be real, 
these terms are purely imaginary and decouple from the real parts  $T^\gamma_{(r)h}$ (equal to the matter tensor) and build 
16 homogeneous (non-linear, second order), partial differential eqs. $T^\gamma_{(i)h} \stackrel != 0 $, which are then  
{\em additional} and independent compared to the corresponding RMT. In this case, we consequently have a set of 32 independent 
real eqs.\FN{
  again look at the linear PPN example in sec \ref{ssec-lin-pn}, where actually all 32 eqs. are solved with real tetrads.
} for {\em only 16 real} tetrad components, which is expected to have  generally no solution.\\
However, one remarkable result of the test cases in section \ref{sec-compare} was, that they all actually {\em can be 
solved with real tetrads} (for the PPN-tests they are real up to the required approximation order).\FN{
  Obviously all statements in this section stay true, if we consider a constant, unitary transformation of all tetrads 
  $e^a_\mu \to e^{i\VP} e^a_\mu$, which also does not affect the metric.
} 
Therefore we shortly list  the general form of these additional conditions  in the following,
although we are not yet able to give a complete mathematical and physical analysis of the solvability with real tetrads.

If we consider only real tetrads, all terms $r^a_{bc}, r_c, t^a, |e|, \dots$ are also real, and the imaginary part of the symbol $U^{[ab]}_x$ 
of eq. \refeq{eq-U-Lv-c}, which builds $T^\gamma_{(i)h}$ becomes
\BE
  U^{[ab]}_{(i) x} = i (\delta^a_x t^{b} - \delta^b_x t^{a}) + i \Delta^{cfab} \eta_{x f}  r_c,\MBOX{i.e.}
  U^{[xb]}_{(i) x} = 3i t^{b} \MBOX{and}  \LAGR_i = 0.
\EE 
If we define the term $T^m_{(i)h} = e^m_\gamma T^\gamma_{(i)h} $, the following 16 conditions result, after some formula manipulations: 
\BE \label{eq-Tmh-real}
 T^m_{(i)h}  = i[(r^a_{{ h}f} \Delta^{cxf{ m}} \eta_{xa}  + \frac 12 r^{ m}_{ab} \Delta^{cfab} \eta_{{ h} f} ) r_c  
  - e^\alpha_b\Delta^{cf{ m}b} \eta_{{ h}f}r_{c,\alpha} + e^\alpha_{ h} t^{ m}_{,\alpha}] \stackrel != 0
\EE
By contracting with $e^h_\beta$ it is possible to derive explicit formulas for $t^m_{,\beta} = \cdots$.
For the trace results the simple divergence-eq.
\BE T^m_{(i)m} = i(-t^m r_m +  e^\alpha_m t^m_{,\alpha}) = \frac i{|e|}   (|e| e^\alpha_m t^m)_{,\alpha} \stackrel != 0
\EE
For the important linear case $e^a_\mu \approx \delta^a_\mu$ we receive the following approximation, which can be expressed
by an antisymmetric ``superpotential'' $F^{mb}_{h} = -F^{bm}_{h}$:\FN{ 
  using  $ T^m_{(i)h}  \approx  -U^{[mb]}_{(i) h,b} =-  i (\Delta^{cfmb} \eta_{h f}  r_{c,b} - t^{m}_{,h}) $
} 
\BE \label{eq-Tmh-real-lin}
 F^{mb}_{h} \DEF -i \Delta^{mbcf}( \eta_{fh} e^a_{c,a} + \eta_{fa} e^a_{c,h}) \MBOX{with}  T^m_{(i)h} = F^{mb}_{h,b} \stackrel != 0.
\EE
The trace vanishes identically $ T^m_{(i)m} \equiv 0$, since $t^m_{,m} \equiv 0$. 
For the solvability of this system it is also important, that the identity $F^{mb}_{h,bm} \equiv 0 $ holds (because of the antisymmetry
of $F$).\\
It remains to clarify, for which matter tensors the eqs. \refeq{eq-Tmh-real} resp. \refeq{eq-Tmh-real-lin} have no solutions, 
which also satisfy \refeq{eq-def-tgh}, i.e.
complex tetrads are actually required.
 
\section{Conclusions and outlook}
Here is  presented a new classical theory of gravitation, which is in most test cases 
(\SC{Schwarzschild}-metric, post-Newtonian approximation),  
identical to the \SC{Einstein}-theory. But unlike other tetrad gravity theories, it does not exhibit
some typical  physical unreasonable vacuum solutions.

It remains to clarify, if the correspondence of the symmetry groups of matrix-theory and standard electro-weak
theory in particle physics $ SL(2) \times U(1) \supset SU(2) \times U(1) $ is merely a pure coincidence, 
or if there are deeper connections between both.
If the latter is the case, this would surely be worth of discussing in another paper. 
It should be possible to extend the global symmetry to a local one by introducing new gauge fields, likewise 
for the GSW-theory. But this is a quite complicated task, also needing a lot of new ideas.
Also the issue of complex vs. real tetrads, requires further investigations. It should be clarified, in which
cases real tetrad solutions are possible and how to interpret the possible imaginary parts physically.
 
As shown in section \ref{new-vac-sol}, there exists a novel type of vacuum solutions, which 
are not present in the \SC{Einstein}-theory.
Although the sources of the field are not yet identified, these solutions have interesting properties
regarding the galaxy rotation problem.
To describe the sources of these solutions, it might be necessary to consider  non-symmetrical
stress-energy matter tensors.
\SC{Einstein} spent his last
years searching for a non-symmetrical field theory \cite{EINSTEIN}, which was supposed to incorporate
also electromagnetism, but without success. 
We know nowadays, however, that a classical field theory will not be able to answer all questions, because the 
wave-function in quantum mechanics cannot be regarded as physical field.

Also the cosmological implications of the matrix-theory should be investigated. A very preliminary, first test
with the simplest real tetrads, which produce the usual cosmological Robertson-Walker metrices, gives additional 
imaginary constraints, which force a spatially flat spacetime metric, i.e. $\kappa = 0$. 
According to current astronomical knowledge, the matter density is nearly equal the critical density 
and does not allow the discrimination of $\kappa$, so there is no contradiction.

A remarkable, quite new perspective of the matrix-theory to spacetime geometry are the {\em absolute} matrices e.g. in \refeq{eq-mv-a}. 
These matrices are by definition {\em invariant} under 
all space-time transformations. 

\section{Acknowledgements}
I am indebted to F. W. Hehl for his invaluable advices and very helpful discussions.
Furthermore, I want to express my gratitude  to the referees for their valuable advices to improve this article.
At last, I want to thank J. Kusche for his support and Marina K\"ohler for reviewing the paper.

\begin{appendix}
\section{Appendix}
\subsection{Matrix calculus}
\label{ssec-matrix}
Here we want to list some formulas for matrix calculations, which are needed for the computations 
in sections \ref{sec-matr-gr}
and \ref{sec-lagr-matr}. 
Although quite elementary, they do not appear in most mathematical textbooks.\\

\underline{ a) For quadratic $ n \times n $  matrices $\MV A, \MV B, \dots $ of {\em arbitrary dimension} $n$ holds the following.}\\
$\bullet$ Matrix factors inside the trace can be rotated cyclically\FN{
  The simple proof starts with $\TR{\MV A\MV B} = \TR{\MV B\MV A} $,
  which follows e.g. from the component representation  $\TR{\MV A\MV B} = \sum_{ij} a_{ij}b_{ji}$.
  Due to the associativity of matrix multiplication this can be extended for more than two matrix factors.
}
\BE \label{eq-tr-cyclical}
\TR{\MV A\MV B \MV C \cdots \MV X } = \TR{\MV B \MV C \cdots\MV X \MV A}  = \TR{ \MV C \cdots \MV X \MV A \MV B} = \cdots . 
\EE
$\bullet$ The trace of a hermitian matrix $\HC {\MV A} = \MV A$ is always a real number $\TR {\MV A }= real$, and also of 
the product of two hermitian matrices  $\TR {\MV A \MV B }= real $ (using eq. \refeq{eq-tr-cyclical}).
But this generally does {\em not} hold for traces of more than two factors.\\
$\bullet$ For the variation principle we need the following theorem:\\ 
The vanishing of the trace $\TR{\VAR \MV x \MV T} \stackrel != 0$
for every variation matrix $\VAR \MV x $ forces the matrix eq. $\MV T \stackrel != 0$.\\

\underline{b) The rest of this section holds for {\em $2\times 2$ matrices only}.}\\
We define for a matrix $\MV A = {\alpha,\beta\choose \gamma,\delta }$ a "bar''-operation ("adjunction") as
the linear map $\bar \MV A \DEF {\;\; \delta,-\beta\choose -\gamma,\;\; \alpha}$.\\
$\bullet$ It is obviously interchangeable with hermitian adjugation 
$\HC {(\bar \MV A)} = \overline{(\HC {\MV A})} $, 
fulfills $\bar {\bar \MV A} = \MV A $ and the evident equations with the identity matrix $\MV I$:
\BE \label{eq-bar-OP-eqs}
 \overline{\MV A \MV B} =\bar \MV B \bar \MV A, \qquad 
\bar \MV A \MV A =\MV A  \bar \MV A = | \MV A| \MV I,\qquad \MV A + \bar \MV A = \TR{\MV A} \MV I,\qquad 
\TR{\MV A} = \TR{\bar \MV A}. 
\EE
$\bullet$ The product of two matrices $\MV A \MV B$ obeys no definite transformation rule under $T$-transformations 
defined in \refeq{eq-T-trafo}, but the ``bar-alternating'' product  $\MV A \bar \MV B$ transforms in a definite manner as
\BE \MV A \bar \MV B \to T \MV A \HC T \; \bar \HC T \bar \MV B \bar T = T (\MV A \bar \MV B) \bar T.
\EE 
The same holds for products of more than 2 matrices.\\
$\bullet$ As a special case of above, the trace of a bar-alternating matrix product with even number of factors is invariant under 
$T$-transformations, e.g.
\BE \TR {\dots \MV A \bar \MV B \MV C \bar \MV D \dots} = inv.
\EE
$\bullet$ If $\MV {x,y,z,u}$ are hermitian matrices, representing Minkowski spacetime vectors, the expressions
\BE \MV F\MV{(x,y)} \DEF \frac i2 (\MV { x\bar y - y\bar x}),\quad \MV V \MV{(x,y,z)} \DEF \frac i2 (\MV{x\bar y z - z\bar y x}), 
\quad V_4(\MV {x,y,z,u}) \DEF \frac 12 \Im \TR{\MV {x\bar y z \bar u}}
\EE
are: $\MV {F(x,y)}$ = {\em area} (non-hermitian, traceless, 6 real comp.),  $\MV {V(x,y,z)}$ = {\em 3-volume} (hermitian, 4 real comp.)
and $V_4(\MV {x,y,z,u})$ =  {\em 4-volume} (real scalar), respectively. 
All three expressions change the sign on odd permutations and vanish for linearly dependent vectors.\\

\underline{c) Relations including the base-matrices $\tau_\mu$:}\\
$\bullet$ For every matrix $\MV A$ hold the three identities (to derive from the orthogonality and completeness of the basis)
\BE \TR{\MV A\bar \tau_\mu} \tau^\mu = \TR{\MV A\bar \tau^\mu} \tau_\mu = 2\MV A, \quad 
  \tau^\mu \bar \MV A \tau_\mu = -2 \MV A,\quad \bar \tau^\mu \MV A \tau_\mu = 2 \MV I \TR{\MV A} = 2 (\MV A + \bar \MV A) .
\EE
$\bullet$ For any non-singular basis ($|\tau|\ne 0$) and any index-combination $\alpha,\beta,\gamma$ holds
\BE 
  \tau^\alpha  \bar \tau^\beta \tau^\gamma - \tau^\gamma  \bar \tau^\beta \tau^\alpha = -2i \epsilon^{\alpha\beta\gamma\lambda} \tau_\lambda  
  \MBOX{and} \epsilon_{\alpha\beta\gamma\lambda}\tau^\alpha  \bar \tau^\beta \tau^\gamma = 6i \tau_\lambda
\EE
where $\epsilon$ is the completely antisymmetric tensor, with the scalar components 
$\epsilon^{0123} = \frac 1{|\tau|},\dots $ and\\ $ \epsilon_{0123} = -|\tau|, \dots$. These formulas allow an explicit
computation of the contravariant- from the covariant matrices and vice versa.\\
$\bullet$ To compute traces of products of {\sc Pauli}-matrices, like in the eq. \refeq{eq-lagr-tetr} 
an  ``index shifting'' technique can be used, which is shortly sketched here. 
It is based on the orthogonality relations eq. \refeq{eq-g-matr-def}, which can also be written as 
$\sigma_m \bar\sigma_l + \sigma_l \bar\sigma_m = 2 \eta_{ml} \MV I$.
We get e.g.
\BE \TR {\sigma_m \bar\sigma_l \sigma^a\cdots } = \TR {(2\eta_{ml} - \sigma_l \bar\sigma_m) \sigma^a\cdots } = 
2\eta_{ml} \TR{\sigma^a\cdots} - \TR{ \sigma_l \bar\sigma_m \sigma^a\cdots}  = \cdots .
\EE
Using this technique multiple times, in combination with the symmetry relations eqs. \refeq{eq-tr-cyclical} and \refeq{eq-bar-OP-eqs}
gives the requested formulas.
One example with 4 {\sc Pauli}-matrices is the identity
\BE \label{eq-4-sigma-tr}
  \frac 12 \TR{\sigma^a \bar \sigma^b \sigma^c \bar \sigma^d } =  
  (\eta^{ab}\eta^{cd} - \eta^{ac}\eta^{bd} + \eta^{ad} \eta^{bc}) - i \Delta^{abcd},
\EE
where $ \Delta^{abcd}$ is the completely antisymmetric symbol, with $\Delta^{0123} = 1$.\\
\subsection{Some explicit {\sc Lagrangian} terms expressed by the symbols $r^a_{bc}$}
\label{ssec-append-lagr}
The following explicit expressions are included, to allow readers to check some formulas in this paper. 
They are computed with the help of a small computer program for symbolic computations ``Symbolic'' \cite{Symbolic}
(see page \pageref{page-symbolic}), but can be easily verified by hand.
For uniqueness, the antisymmetric $r^a_{bc}$ are always selected by the index combination $b < c$.
Then the contracted terms of eq. \refeq{eq-def-rt} are explicitly given as
\BE 
r_0 =   r^1_{0 1} +  r^2_{0 2} +  r^3_{0 3}, \quad
r_1 =  -  r^0_{0 1} +  r^2_{1 2} +  r^3_{1 3}, \quad
r_2 =  -  r^0_{0 2} -  r^1_{1 2} +  r^3_{2 3}, \quad
r_3 =  -  r^0_{0 3} -  r^1_{1 3} -  r^2_{2 3} 
\EE
\BE
t^0  =   -  r^1_{2 3} +  r^2_{1 3} -  r^3_{1 2}, \quad
t^1  =   -  r^0_{2 3} -  r^2_{0 3} +  r^3_{0 2}, \quad
t^2  =    r^0_{1 3} +  r^1_{0 3} -  r^3_{0 1}, \quad
t^3  =   -  r^0_{1 2} -  r^1_{0 2} +  r^2_{0 1}
\EE

First we list some {\bf terms of the general Lagrangian} in eq. \refeq{eq-Lag-abcde}. 
\begin{eqnarray} 
  \LAGR_a  & \DEF &  \eta^{m n} r_{m} r_{n}^* =   \nonumber
   - (r^{0}_{01} - r^{2}_{12} - r^{3}_{13}) ( r^{0 *}_{01} -  r^{2 *}_{12} - r^{3 *}_{13})  
  - (  r^{0}_{02} + r^{1}_{12} - r^{3}_{23})  ( r^{0 *}_{02}+ r^{1 *}_{12} -  r^{3 *}_{23}) 
  \\&& 
  - ( r^{0}_{03} + r^{1}_{13} + r^{2}_{23})  ( r^{0 *}_{03} +  r^{1 *}_{13} +  r^{2 *}_{23}  )
  + (  r^{1}_{01} + r^{2}_{02} + r^{3}_{03})  ( r^{1 *}_{01} + r^{2 *}_{02} + r^{3 *}_{03})
\\   \LAGR_b &\DEF & \eta^{m n} r^{a}_{m b} (r^{b }_{n a})^* =
  - r^{0}_{01} r^{0 *}_{01} - r^{0}_{02} r^{0 *}_{02} - r^{0}_{03} r^{0 *}_{03}+ (  r^{0}_{12} + r^{1}_{02})  r^{2 *}_{01}+ 
  (  - r^{0}_{12} + r^{2}_{01})  r^{1 *}_{02}+ (  r^{0}_{13} 
  + r^{1}_{03})  r^{3 *}_{01} \nonumber \\ &&  + (  - r^{0}_{13} + r^{3}_{01})  r^{1 *}_{03}+ (  r^{0}_{23} + r^{2}_{03})  r^{3 *}_{02}+ 
  (  - r^{0}_{23} + r^{3}_{02})  r^{2 *}_{03} + r^{1}_{01} r^{1 *}_{01}
  + (  - r^{1}_{02} + r^{2}_{01})  r^{0 *}_{12}   \\ &&  + (  - r^{1}_{03} + r^{3}_{01})  r^{0 *}_{13} - r^{1}_{12} r^{1 *}_{12} - r^{1}_{13} r^{1 *}_{13}+ 
  (  r^{1}_{23} - r^{2}_{13})  r^{3 *}_{12}
  + (  - r^{1}_{23} - r^{3}_{12})  r^{2 *}_{13} + r^{2}_{02} r^{2 *}_{02}  \nonumber \\ &&   + (  - r^{2}_{03} + r^{3}_{02})  r^{0 *}_{23} - r^{2}_{12} r^{2 *}_{12}+ 
  (  - r^{2}_{13} + r^{3}_{12})  r^{1 *}_{23} 
   - r^{2}_{23} r^{2 *}_{23} + r^{3}_{03} r^{3 *}_{03} - r^{3}_{13} r^{3 *}_{13} - r^{3}_{23} r^{3 *}_{23}
  \nonumber \\ 
  \LAGR_c &\DEF &\eta^{m n} \eta_{a b} \eta^{c d} r^{a}_{m c} (r^{b}_{n d})^* =  
  -2 r^{0}_{01} r^{0 *}_{01} -2 r^{0}_{02} r^{0 *}_{02} -2 r^{0}_{03} r^{0 *}_{03} + 2 r^{0}_{12} r^{0 *}_{12} + 2 r^{0}_{13} r^{0 *}_{13} + 
  2 r^{0}_{23} r^{0 *}_{23} + 2 r^{1}_{01} r^{1 *}_{01} \nonumber \\ &&  
  + 2 r^{1}_{02} r^{1 *}_{02} + 2 r^{1}_{03} r^{1 *}_{03} -2 r^{1}_{12} r^{1 *}_{12} -2 r^{1}_{13} r^{1 *}_{13} -2 r^{1}_{23} r^{1 *}_{23} + 
  2 r^{2}_{01} r^{2 *}_{01} + 2 r^{2}_{02} r^{2 *}_{02} 
  + 2 r^{2}_{03} r^{2 *}_{03} -2 r^{2}_{12} r^{2 *}_{12}  \nonumber \\ &&   - 2 r^{2}_{13} r^{2 *}_{13} -2 r^{2}_{23} r^{2 *}_{23} + 2 r^{3}_{01} r^{3 *}_{01} + 
  2 r^{3}_{02} r^{3 *}_{02} + 2 r^{3}_{03} r^{3 *}_{03} 
  -2 r^{3}_{12} r^{3 *}_{12} -2 r^{3}_{13} r^{3 *}_{13} -2 r^{3}_{23} r^{3 *}_{23}
  \\ \LAGR_x &\DEF & \LAGR_c - 2\LAGR_b = -2\eta_{ab} t^a t^{b*} =
  2(   r^{0}_{12} +  r^{1}_{02} - r^{2}_{01})  (r^{0 *}_{12}+  r^{1 *}_{02} -  r^{2 *}_{01} )
  + 2(  r^{0}_{13} +  r^{1}_{03} - r^{3}_{01}) ( r^{0 *}_{13}+  r^{1 *}_{03} - r^{3 *}_{01} ) \nonumber \\ &&  
  + 2( r^{0}_{23} +  r^{2}_{03} - r^{3}_{02}) ( r^{0 *}_{23}+  r^{2 *}_{03} - r^{3 *}_{02}) 
  - 2(  r^{1}_{23} -  r^{2}_{13} + r^{3}_{12})  ( r^{1 *}_{23} - r^{2 *}_{13} + r^{3 *}_{12} )
\end{eqnarray} 
The {\sc Einstein-Lagrangian} $\LAGR_E$ reads explicitly (for real tetrads):
\begin{eqnarray} 
  \LAGR_E &=& \LAGR_a - \frac 12 \LAGR_b - \frac 14 \LAGR_c =
  2 r^{0}_{01} r^{2}_{12}+ 2( r^{0}_{01} - r^{2}_{12})  r^{3}_{13} -2 r^{0}_{02} r^{1}_{12}+ 2(r^{0}_{02} + r^{1}_{12})  r^{3}_{23}-2 r^{0}_{03} r^{1}_{13}
  \\ && \nonumber  
  -2 (r^{0}_{03} + r^{1}_{13})  r^{2}_{23} -\frac 12 (r^{0}_{12})^2 -\frac 12 (r^{0}_{13})^2 -\frac 12 (r^{0}_{23})^2 + 2 r^{1}_{01} r^{2}_{02}+ 
  2( r^{1}_{01} + r^{2}_{02})  r^{3}_{03}
  + r^{0}_{12} r^{1}_{02} 
  \\ \nonumber &&  -\frac 12 (r^{1}_{02})^2 + r^{0}_{13} r^{1}_{03} -\frac 12 (r^{1}_{03})^2 + \frac 12 (r^{1}_{23})^2
  - ( r^{1}_{02} + r^{0}_{12})  r^{2}_{01} 
  -\frac 12 (r^{2}_{01})^2 + r^{0}_{23} r^{2}_{03}  -\frac 12 (r^{2}_{03})^2 
  \\  \nonumber && + r^{1}_{23} r^{2}_{13} + \frac 12 (r^{2}_{13})^2
  -( r^{1}_{03} +r^{0}_{13})  r^{3}_{01} -\frac 12 (r^{3}_{01})^2 
  -( r^{2}_{03} + r^{0}_{23})  r^{3}_{02} -\frac 12 (r^{3}_{02})^2+ (  r^{2}_{13} - r^{1}_{23})  r^{3}_{12}  + \frac 12 (r^{3}_{12})^2
\end{eqnarray} 
The two terms of the {\bf matrix} {\sc Lagrangian} $\LAGR_z = \LAGR_r + i\LAGR_i$ in eq. \refeq{eq-lagr_z} are
\begin{eqnarray} 
  \LAGR_r & = & \LAGR_a -\LAGR_b =
  (  r^{2}_{12} + r^{3}_{13})  r^{0 *}_{01}+ (  - r^{1}_{12} + r^{3}_{23})  r^{0 *}_{02}+ (  - r^{1}_{13} - r^{2}_{23})  r^{0 *}_{03}+ 
  (  r^{1}_{02} - r^{2}_{01})  r^{0 *}_{12} \\ &&  
  + (  r^{1}_{03} - r^{3}_{01})  r^{0 *}_{13}+ (  r^{2}_{03} - r^{3}_{02})  r^{0 *}_{23}+ (  r^{2}_{02} + r^{3}_{03})  r^{1 *}_{01}+ 
  (  r^{0}_{12} - r^{2}_{01})  r^{1 *}_{02}  \nonumber \\ &&  
  + (  r^{0}_{13} - r^{3}_{01})  r^{1 *}_{03}+ (  - r^{0}_{02} + r^{3}_{23})  r^{1 *}_{12}+ (  - r^{0}_{03} - r^{2}_{23})  r^{1 *}_{13}+ 
  (  r^{2}_{13} - r^{3}_{12})  r^{1 *}_{23}  \nonumber \\ &&  
  + (  - r^{0}_{12} - r^{1}_{02})  r^{2 *}_{01}+ (  r^{1}_{01} + r^{3}_{03})  r^{2 *}_{02}+ (  r^{0}_{23} - r^{3}_{02})  r^{2 *}_{03}+ 
  (  r^{0}_{01} - r^{3}_{13})  r^{2 *}_{12}  \nonumber \\ &&  
  + (  r^{1}_{23} + r^{3}_{12})  r^{2 *}_{13}+ (  - r^{0}_{03} - r^{1}_{13})  r^{2 *}_{23}+ (  - r^{0}_{13} - r^{1}_{03})  r^{3 *}_{01}+ 
  (  - r^{0}_{23} - r^{2}_{03})  r^{3 *}_{02}  \nonumber \\ &&  
  + (  r^{1}_{01} + r^{2}_{02})  r^{3 *}_{03}+ (  - r^{1}_{23} + r^{2}_{13})  r^{3 *}_{12}+ (  r^{0}_{01} - r^{2}_{12})  r^{3 *}_{13}+ 
  (  r^{0}_{02} + r^{1}_{12})  r^{3 *}_{23}
  \nonumber\\ \LAGR_i &=& 
  (  r^{0}_{23} + r^{2}_{03} - r^{3}_{02})  (r^{0 *}_{01} -   r^{2 *}_{12} - r^{3 *}_{13}) 
  + (  - r^{0}_{13} - r^{1}_{03} + r^{3}_{01})  (r^{0 *}_{02} + r^{1 *}_{12} - r^{3 *}_{23})
  \\&&   + (  - r^{1}_{23} + r^{2}_{13} - r^{3}_{12})  ( r^{1 *}_{01} +  r^{2 *}_{02} + r^{3 *}_{03})
  + (  r^{0}_{12} + r^{1}_{02} - r^{2}_{01})  (r^{0 *}_{03}  +  r^{1 *}_{13} +  r^{2 *}_{23} ) 
 \nonumber   \\ &&  
  + (  - r^{1}_{13} - r^{2}_{23} - r^{0}_{03})  (r^{0 *}_{12} +  r^{1 *}_{02} - r^{2 *}_{01})
  + (  r^{1}_{12} - r^{3}_{23} + r^{0}_{02})  ( r^{0 *}_{13} + r^{1 *}_{03} -  r^{3 *}_{01})
  \nonumber  \\&&  + (  r^{2}_{12} + r^{3}_{13} - r^{0}_{01})  (r^{0 *}_{23}  +  r^{2 *}_{03} - r^{3 *}_{02}) 
  +    (  r^{2}_{02} + r^{3}_{03} + r^{1}_{01})  ( r^{1 *}_{23} - r^{2 *}_{13} + r^{3 *}_{12})  \nonumber
\end{eqnarray} 
$\LAGR_a$ consists solely of ``r-doublets'', $\LAGR_x$ solely of ``r-triplets''. None of the Lagrangians $\LAGR_r,\LAGR_i,\LAGR_E $ 
contains quadrats of r-doublets. 
In the terms $\LAGR_a,\LAGR_b, \LAGR_c, \LAGR_r, \LAGR_E $ r-doublets and r-triplets do not mix, while $\LAGR_i$ consists solely 
of mixed products. $\LAGR_z = \LAGR_r +i \LAGR_i$ does not contain quadrats of r-triplets.
All $\LAGR_a,\LAGR_b, \LAGR_c$ have even parity and only $\LAGR_i$ has odd parity.\\
The {\bf generalized ``viable''} {\sc Lagrangian} of section \ref{sec-iso-sm}  has the form $\LAGR_v(c) = \LAGR_z + c(\LAGR_c -2 \LAGR_b) =  \LAGR_z + c\LAGR_x$
(definition in eq. \refeq{eq-lagr_v}, but only for the special case $d = -1/2$, see footnote \ref{fn-omit-d}).
The antisymmetrized $U$-terms, defined in  eq. \refeq{eq-def-ufba} and \refeq{eq-U-Lv-c}, for  this $\LAGR_v(c)$ 
are explictly (we list here 6 representatives, the other 18 symbols are similar)
\begin{eqnarray}
  \label{eq-U-Lv}
  U_{0}^{[0 1]} &=& - i r^{0*}_{23} - i r^{2*}_{03} + i r^{3*}_{02} + r^{2*}_{12} + r^{3*}_{13}
  \\ U_{0}^{[1 2]} &=&  -(  2 c +1)  r^{2*}_{01}+ (  2 c + 1)  r^{1*}_{02} + 2 c r^{0*}_{12} + i r^{0*}_{03} + i r^{1*}_{13} + i r^{2*}_{23}
  \nonumber  \\U_{1}^{[0 1]} &=&   i r^{1*}_{23} - i r^{2*}_{13} + i r^{3*}_{12} + r^{2*}_{02} + r^{3*}_{03}
  \nonumber  \\U_{1}^{[0 2]} &=&  -(2 c +1)  r^{2*}_{01}+ (  2 c + 1)  r^{0*}_{12} + 2 c r^{1*}_{02} + i r^{0*}_{03} + i r^{1*}_{13} + i r^{2*}_{23}
  \nonumber  \\U_{1}^{[1 2]} &=&  i r^{0*}_{13} + i r^{1*}_{03} - i r^{3*}_{01} - r^{0*}_{02} + r^{3*}_{23}
  \nonumber \\U_{1}^{[2 3]} &=& -(2 c +1)  r^{3*}_{12}+ (  2 c + 1)  r^{2*}_{13} -2 c r^{1*}_{23} - i r^{1*}_{01} - i r^{2*}_{02} - i r^{3*}_{03}
  \nonumber
\end{eqnarray} 

\subsection{Computation of \SC{Riemann}-, \SC{Ricci}-tensors and $R$ with tetrads}
\label{ssec-append-1}
The aim of this section is to compute the $R$ scalar with the tetrad-formalism of section \ref{sec-matr-gr}
to enable its comparison with the \SC{Lagrangian} of the matrix-theory, as presented in section  \ref{sec-gen-lagr}.\\
The \SC{Riemann}-tensor is defined as the $[\lambda\nu]$-antisymmetric expression
\BE \label{eq-Riemann-def}
  R^\sigma_{\;\;\mu\nu\lambda} \DEF 
  \underbrace{\Gamma^\sigma_{\mu\lambda,\nu}   + \Gamma^\sigma_{\alpha\nu}\Gamma^\alpha_{\mu\lambda}}_{
    \DEF S^\sigma_{\;\;\mu\lambda\nu} }
  - \underbrace{\Gamma^\sigma_{\mu\nu,\lambda} -\Gamma^\sigma_{\alpha\lambda}\Gamma^\alpha_{\mu\nu}}_{
    \DEF S^\sigma_{\;\;\mu\nu\lambda} }
  \quad =  S^\sigma_{\;\;\mu\lambda\nu} -  S^\sigma_{\;\;\mu\nu\lambda} = S^\sigma_{\;\;\mu[\lambda\nu]}.
\EE
The \SC{Christoffel}-symbols therein can be expressed by the tetrads (we only consider {\em real } tetrads here,
because they suffice to describe  \SC{Riemann}-spacetime)
using the standard formula
 \BEA
  \Gamma^\sigma_{\mu\nu} &=& 
  \frac 12 g^{\sigma\alpha}(g_{\mu\alpha,\nu} +  g_{\nu\alpha,\mu} - g_{\mu\nu,\alpha})\\
  &= &  \frac 12 g^{\sigma\alpha}( (e^a_\mu e_{a\alpha})_{,\nu} + (e^a_\nu e_{a\alpha})_{,\mu}  
  - (e^a_\mu  e_{a\nu})_{,\alpha}) \\
  &=& \frac 12 g^{\sigma\alpha} ( e_{c\mu} (e^c_{\alpha,\nu} -e^c_{\nu,\alpha} )
  +  e_{c\nu} (e^c_{\alpha,\mu} -e^c_{\mu,\alpha}) 
  + e_{c\alpha} (e^c_{\mu,\nu} + e^c_{\nu,\mu}) )\\
  &=& \frac 12 g^{\sigma\alpha} ( e_{c\mu} e^c_{[\alpha,\nu]} +  e_{c\nu} e^c_{[\alpha,\mu]}) 
  + \frac 12 e_c^\sigma e^c_{(\mu,\nu)} .
\EEA
We introduce the new symbols $\Gamma^s_{\mu\nu}$ by transforming the upper index
into tetrad type $\sigma \to s$
\BE 
   \Gamma^s_{\mu\nu} \DEF e^s_\sigma \Gamma^\sigma_{\mu\nu} \quad  \lra \quad
   e^\sigma_s \Gamma^s_{\mu\nu} = \Gamma^\sigma_{\mu\nu} ,
\EE
and with them the covariant tetrad derivative is defined as the expression 
(in contrast to the $\Gamma^s$-symbols, the $\GT^s$ are obviously tensors):
\BEN \label{eq-tetr-cov-der}
 \GT^s_{\mu\nu}  &\DEF&  e^s_{\mu;\nu} = e^s_{\mu,\nu} - \Gamma^\sigma_{\mu\nu} e^s_\sigma = 
  e^s_{\mu,\nu} - \Gamma^s_{\mu\nu}\\
  &=&  \frac 12 (e^s_{[\mu,\nu]} + e^{s\alpha}( e_{c\mu} e^c_{[\nu,\alpha]} +  e_{c\nu} e^c_{[\mu,\alpha ]})).
\EEN
In the following we also will need their tetrad components, which are with the definitions in eq. \refeq{eq-def-rafb} 
\BE \label{eq-ricci-cr}
  \GT^s_{mn}  \DEF  e^\mu_m e^\nu_n \GT^s_{\mu\nu} =
  \frac 12 (r^s_{mn} + \eta^{sb}(\eta_{mc} r^c_{nb} +\eta_{nc} r^c_{mb})) .
\EE
In some references these termes, which are by definition scalars, 
are titled ``Ricci's coefficients of rotation''.
Then we can compute the second summand as\FN{
  The derivatives of contravariant tetrads are obtained from the orthogonality relations as
  $ e^\sigma_{s,\lambda} = -e^\alpha_s e^\sigma_b e^b_{\alpha,\lambda}$.
}
\BEN 
  S^{ \sigma }_{\;\;\mu\nu\lambda} &=&
  \Gamma^\sigma_{\mu\nu,\lambda}  + \Gamma^\sigma_{\alpha\lambda}\Gamma^\alpha_{\mu\nu}\\
  &=&  (e^\sigma_s \Gamma^s_{\mu\nu})_{,\lambda}  + \Gamma^\sigma_{\alpha\lambda}e^\alpha_s\Gamma^s_{\mu\nu}
  =  e^\sigma_s \Gamma^s_{\mu\nu,\lambda}  + 
  \Gamma^s_{\mu\nu}( e^\sigma_{s,\lambda}  + e^\alpha_s\Gamma^\sigma_{\alpha\lambda})\\
  &=&    e^\sigma_s \Gamma^s_{\mu\nu,\lambda}  +   \Gamma^s_{\mu\nu}(  -e^\alpha_s e^\sigma_b e^b_{\alpha,\lambda}  
  + e^\alpha_s e^\sigma_b\Gamma^b_{\alpha\lambda})\\
 &=&    e^\sigma_s \Gamma^s_{\mu\nu,\lambda}  -   
  e^\alpha_s e^\sigma_b \Gamma^s_{\mu\nu}(e^b_{\alpha,\lambda}  -\Gamma^b_{\alpha\lambda}) =
    e^\sigma_s (\Gamma^s_{\mu\nu,\lambda}  -  \Gamma^\alpha_{\mu\nu} \GT^s_{\alpha\lambda} ) ,
\EEN
and we get\FN{
  with $e^s_{(\mu,\nu),\lambda} - e^s_{(\mu,\lambda),\nu} = 
  (e^s_{\mu,\nu}+ e^s_{\nu,\mu})_{,\lambda} - (e^s_{\mu,\lambda} + e^s_{\lambda,\mu})_{,\nu} =
  e^s_{\nu,\mu\lambda} -   e^s_{\lambda,\mu\nu} =     e^s_{[\nu,\lambda],\mu} =   
  - e^s_{[\mu,\nu],\lambda} +e^s_{[\mu,\lambda],\nu}   $
}
\BEN \label{eq-tetr-R-sigma}
  R^\sigma_{\;\;\mu\nu\lambda} &=&   S^\sigma_{\;\;\mu\lambda\nu} -  S^\sigma_{\;\;\mu\nu\lambda}\\ 
  &=&  e^\sigma_s (\underbrace{\Gamma^s_{\mu\lambda,\nu} -\Gamma^s_{\mu\nu,\lambda}  }_{  
    = \GT^s_{\mu\nu,\lambda} -\GT^s_{\mu\lambda,\nu}  } 
  +\Gamma^\alpha_{\mu\nu} \GT^s_{\alpha\lambda} -\Gamma^\alpha_{\mu\lambda} \GT^s_{\alpha\nu}) \nonumber\\
  &=&  e^\sigma_s (\GT^s_{\mu\nu,\lambda} -\GT^s_{\mu\lambda,\nu}  
  + \Gamma^\alpha_{\mu\nu} \GT^s_{\alpha\lambda} -\Gamma^\alpha_{\mu\lambda} \GT^s_{\alpha\nu})\nonumber
  =  e^\sigma_s  R^s_{\;\;\mu\nu\lambda} .
\EEN
With this we can compute the tetrad components as
\BEA
   R^s_{\;mnl} &=&  e^\mu_m e^\nu_n e^\lambda_l R^s_{\;\;\mu\nu\lambda} = 
  e^\mu_m e^\nu_n e^\lambda_l(\GT^s_{\mu\nu,\lambda} -\GT^s_{\mu\lambda,\nu}  
  + \Gamma^\alpha_{\mu\nu} \GT^s_{\alpha\lambda} -\Gamma^\alpha_{\mu\lambda} \GT^s_{\alpha\nu} )\\
  &=&  e^\mu_m e^\nu_n e^\lambda_l((\GT^s_{xy} e^x_\mu e^y_\nu)_{,\lambda} -(\GT^s_{xy} e^x_\mu e^y_\lambda)_{,\nu}  ) +
  \Gamma^a_{mn}\GT^s_{al} - \Gamma^a_{ml}\GT^s_{an} \\
  &=&  e^\mu_m e^\nu_n e^\lambda_l(\GT^s_{xy,\lambda} e^x_\mu e^y_\nu +\GT^s_{xy} (e^x_\mu e^y_\nu)_{,\lambda} 
  -\GT^s_{xy,\nu} e^x_\mu e^y_\lambda -\GT^s_{xy} (e^x_\mu e^y_\lambda)_{,\nu}  ) 
  + \Gamma^a_{mn}\GT^s_{al} - \Gamma^a_{ml}\GT^s_{an} \\
  &=&  e^\lambda_l \GT^s_{mn,\lambda}  -  e^\nu_n \GT^s_{ml,\nu} 
  + \GT^s_{xy} e^\mu_m e^\nu_n e^\lambda_l ( (e^x_\mu e^y_\nu)_{,\lambda} - (e^x_\mu e^y_\lambda)_{,\nu}  )
  + \Gamma^a_{mn}\GT^s_{al} - \Gamma^a_{ml}\GT^s_{an} \\
  &=&  e^\lambda_l \GT^s_{mn,\lambda} - e^\lambda_n\GT^s_{ml,\lambda}
  + \GT^s_{xy} e^\mu_m e^\nu_n e^\lambda_l 
  ( e^x_\mu \underbrace{(e^y_{\nu,\lambda} - e^y_{\lambda,\nu} )}_{e^y_{[\nu,\lambda]}} + 
   e^y_\nu  e^x_{\mu,\lambda}-  e^y_\lambda e^x_{\mu,\nu})
  + \Gamma^a_{mn}\GT^s_{al} - \Gamma^a_{ml}\GT^s_{an} \\
  &=& e^\lambda_p (\delta^p_l \GT^s_{mn} - \delta^p_n \GT^s_{ml} )_{,\lambda}
  + \GT^s_{xy}( \delta^x_m r^y_{nl} +  \delta^y_n   e^x_{ml}-  \delta^y_l e^x_{mn})
  - \GT^s_{an} \Gamma^a_{ml} + \GT^s_{al} \Gamma^a_{mn} \\
  &=& e^\lambda_p (\delta^p_l \GT^s_{mn} - \delta^p_n \GT^s_{ml} )_{,\lambda}
  + \GT^s_{xy}( \delta^x_m r^y_{nl} + \delta^y_n  e^x_{ml}  - \delta^y_l e^x_{mn} - \delta^y_n \Gamma^x_{ml}
  + \delta^y_l\Gamma^x_{mn})\\
  &=&  e^\lambda_p (\delta^p_l \GT^s_{mn} - \delta^p_n \GT^s_{ml} )_{,\lambda}
  + \GT^s_{xy}( \delta^x_m r^y_{nl} + \delta^y_n  (\underbrace{e^x_{ml} - \Gamma^x_{ml}}_{ = \GT^x_{ml}}) 
  - \delta^y_l (\underbrace{e^x_{mn} -\Gamma^x_{mn}}_{= \GT^x_{mn}})),
\EEA
so we have finally the tetrad representation of the \SC{Riemann} tensor: 
\BE \label{eq-tetr-R-smnl}
    \underline{ R^s_{\;mnl}=  e^\lambda_p (\delta^p_l \GT^s_{mn} - \delta^p_n \GT^s_{ml} )_{,\lambda}
      + \GT^s_{xy}( \delta^x_m r^y_{nl} + \delta^y_n \GT^x_{ml} - \delta^y_l\GT^x_{mn})}.
\EE
Remarkable  in this representation is the fact, that it is completely expressed by the $\GT$ and thus the $r$-terms, which
in turn can be expressed by the $\rho$-tensor-matrix.

From this we get by contracting over first and fourth index the tetrad components of the \SC{Ricci} tensor as
\BEN \label{eq-tetr-Ricci-mn}
   R_{mn} &\DEF &  R^s_{\;mns}\\
   &=&  e^\lambda_p (\delta^p_s \GT^s_{mn} - \delta^p_n \GT^s_{ms} )_{,\lambda}
   + \GT^s_{xy}( \delta^x_m r^y_{ns} + \delta^y_n \GT^x_{ms} - \delta^y_s\GT^x_{mn})\nonumber\\
   &= &  e^\lambda_p (\GT^p_{mn} - \delta^p_n \GT^s_{ms} )_{,\lambda}
   + \GT^s_{xy}( \delta^x_m r^y_{ns} + \delta^y_n \GT^x_{ms} - \delta^y_s\GT^x_{mn}) \nonumber
\EEN
and finally the $R$ scalar\FN{
  The the brackets $[xy],(xy)$ denote the symmetry-type for better readability.
  E.g. the term $ r^c_{ab}\delta^s_c\delta^a_x \delta^b_y \to [xy] $ is antisymmetric.
  Mixed type products $[xy]\times (xy)$ always vanish.
}
\BEA
  R &\DEF&  \eta^{mn} R_{mn} =  
  e^\lambda_p (\eta^{mn} \GT^p_{mn} - \eta^{mp} \GT^s_{ms} )_{,\lambda}
  + \GT^s_{xy}( \eta^{xn} r^y_{ns} + \eta^{ym} \GT^x_{ms} - \delta^y_s \eta^{mn}\GT^x_{mn})\\
  &=&  2 e^\lambda_p \eta^{pb} r^n_{nb,\lambda}
  + \GT^s_{xy}( \eta^{xn} r^y_{ns} + \eta^{ym} \GT^x_{ms}) - r^s_{xs} \eta^{xb}r^n_{nb}\\
  &=&  2 e^{\lambda b} r^n_{nb,\lambda}
  +  \GT^s_{xy}  ( \eta^{xn} r^y_{ns} + \eta^{ym} \frac 12(r^x_{ms}  +   \eta^{xd}
  (\underbrace{\eta_{mc} r^c_{sd}}_{\to r^y_{sd}} + \eta_{sc} r^c_{md}) )) 
  - \underbrace{r^s_{xs}}_{= r_x} \eta^{xb}\underbrace{r^n_{nb}}_{ = -r_b}\\
  &=&  -2 e^{\lambda b} r_{b,\lambda}
  +  \GT^s_{xy}  ( \underbrace{\eta^{xn} r^y_{ns} +  \frac 12 \eta^{xd}  r^y_{sd}} + 
  \frac 12 \eta^{ym} (r^x_{ms}  +  \eta^{xd} \eta_{sc} r^c_{md}) )   + \eta^{xb}r_x r_b\\
  &=&  -2 e^{\lambda b} r_{b,\lambda}
  +  \GT^s_{xy}  \frac 12 (\eta^{xn} r^y_{ns} + 
  \eta^{ym} (r^x_{ms}  +  \eta^{xd} \eta_{sc} r^c_{md}) ) + \eta^{xb}r_x r_b\\
  &=&  -2 e^{\lambda b} r_{b,\lambda} + 
  \frac 14 r^c_{ab}(\underbrace{\delta^s_c\delta^a_x \delta^b_y}_{\to [xy]} 
  +\eta^{sb}\underbrace{(\eta_{xc}\delta^a_y + \eta_{yc} \delta^a_x)}_{(xy)} ) 
  (\underbrace{\eta^{xn} r^y_{ns} + \eta^{yn} r^x_{ns}}_{(xy)}  + 
  \underbrace{\eta^{ym} \eta^{xd} \eta_{sp} r^p_{md}}_{[xy]} ) + \eta^{xb}r_x r_b 
  \\  &=&  -2 e^{\lambda b} r_{b,\lambda} + 
  \frac 14 r^c_{ab} (2\eta^{sb}(r^a_{cs} + \eta_{xc}\eta^{an} r^x_{ns}) + \eta^{bm} \eta^{ad} \eta_{cp} r^p_{md})
  + \eta^{xb}r_x r_b 
  \\  &=&  \underline {-2 e^{\lambda b} r_{b,\lambda}  + 
  \frac 14 r^c_{ab} \eta^{sb} (2r^a_{cs} + \eta_{xc}\eta^{an} r^x_{ns}) + \eta^{xb}r_x r_b} \;.
\EEA

\end{appendix}


\begin{thebibliography}{99}



\newcommand{\SL} {} 
 
\bibitem{ARCOS} Arcos, H.I., Pereira, J.G.: {\SL Torsion and the gravitational  interaction}, 
  Class. Quant. Grav. {\bf 21}, 5193-5202 (2004)

\bibitem{ALDRO} Aldovandi, R., Pereira, J.G., Vu,  K.H.: {\SL Gravitation: global formulation and quantum effects},
 Class. Quant. Grav. {\bf 21}, 51-62 (2004)

\bibitem{BEKENSTEIN} Bekenstein, J.D.: {\SL Revised gravitation theory for the modified Newtonian dynamics paradigm}, 
Phys. Rev. D {\bf 70}, 083509 (2004)
 
\bibitem{BRANS} Brans, C.,  Dicke, R.H.: {\SL Mach's principle and a relativistic theory of gravitation},
Phys. Rev. {\bf 124}, 925-935 (1961)

\bibitem{BROWST-MOFFAT} Brownstein, J.R., Moffat,  J.W.: {\SL Galaxy rotation curves without nonbaryonic
dark matter}, The Astrophys. Journal, 636:721-741, (2006)

\bibitem{CHEN} Chen, H.H., Chern,  D.C., Nester,  J.M.: Chinese J. Phys. {\bf 25}, 481 (1987)

\bibitem{CHENG} Cheng, W.-H., Chern, D.C., Nester,  J.M.: {\SL Canonical analysis of the one-parameter
teleparallel theory}, Phys. Rev. D {\bf 38},  2656 (1988)

\bibitem{EBERT} Ebert, D.: {\SL Eichtheorien (Gauge-theories)}, Akademie Verlag, Berlin (1989)

\bibitem{EINSTEIN} Einstein, A.: {\SL Grundz\"uge der Relativit\"atstheorie}, Akademie Verlag, 
Berlin (1973)

\bibitem{ESTABROOK} Estabrook, F.B.: {\SL Conservation laws for vacuum tetrad gravity}, Class. Quant. Grav., 
{\bf 23}, 2841-2848 (2006)

\bibitem{ACES} European Space Agency (ESA), {\SL Atomic Clock Ensemble in Space (ACES)}\\
  {\tt http://www.spaceflight.esa.int/projects}.

\bibitem{FLECHTNER} Flechtner, F., Neumayer, K., Kusche, J., Sch\"afer, W., Sohl, F.: 
{\SL Simulation study for the determination of the lunar gravity field from PRARE-L
tracking onboard the German LEO mission},  Adv. Space Res. (2007)  (submitted)

\bibitem{HEHL-1996} Gronwald, F., Hehl, F.W.: {On the Gauge Aspects of Gravity},\\
{\tt http://arxiv.org/pdf/gr-qc/9602013} (1996)

\bibitem{GULL} Gull, S., Lasenby, A., Doran, C.: {\SL Imaginary Numbers Are Not Real - The Geometric Algebra of Spacetime},
and subsequent articles, Found. Phys., {\bf 23}, No. 9 (1993)

\bibitem{HAMMOND} Hammond, R.: {\SL Tetrad Formulation of Gravity with a Torsion Potential}, Gen. Rel. Grav., {\bf 26}, No. 11 (1994)

\bibitem{IORIO} Iorio, L.: {\SL Is it possible to test directly general relativity in the
gravitational field of the Moon?}, Class. Quant. Grav. {\bf 19}, 2393-2398 (2002)

\bibitem{ITIN-1}  Itin, Y.: {\SL Coframe Energy-Momentum Current. Algebraic Properties}, Gen. Rel. Grav. {\bf 34}, 1819  (2002)

\bibitem{ITIN-2}  Itin, Y.: {\SL Energy-momentum current for coframe gravity}, Class. Quant. Grav. {\bf 19}, 173  (2002)

\bibitem{KOE-Mink} K\"ohler, W.: {\SL Matrix Representation of Special Relativity},
{\tt http://arxiv.org/pdf/physics/0701105} (2007)


\bibitem{KOPCZ}  Kopczy\'{n}ski, W.: {\SL Problems with metric-teleparallel theories of gravitation}, 
                 J. Phys. A: Math. Gen. {\bf 15}, 493-506 (1982)

\bibitem{KUHF} Kuhfuss, R., Nitsch, J.: {\SL Propagating Modes in Gauge Field Theories of Gravity}, 
               Gen. Rel. Grav. {\bf 18}, 1207  (1986)

\bibitem{KUSCHE} Kusche, J.: {\SL Relativistic Modeling for Geodetic Experiments in Local Spacetimes},  
                 Reihe A, Heft 110, Deutsche Geod\"atische Kommission, M\"unchen (1996)

\bibitem{LL} Landau, Liftschitz, {\SL Quanten-Elektrodynamik}, Akademie Verlag, Berlin (1991)

\bibitem{MALUF-2001} Maluf, J.W., Goya, A.: {\SL Space-time defects and teleparallelism}, 
                     Class. Quant. Grav. {\bf 18}, 5143-5154  (2001)

\bibitem{MALUF} Maluf, J.W., Faria,F.F., Ulhoa, S.C.: {\SL On reference frames in spacetime and gravitational energy
in freely falling frames}, Class. and Quant. Grav., {\bf 24}, 2743-2753  (2007)

\bibitem{MEI} Mei, T.: {\SL A New Variable in General Relativity and Its Applications for Classic and Quantum Gravity},
{\tt http://arxiv.org/pdf/gr-qc/0611063} (2006)

\bibitem{MOLLER} M{\o}ller, C.: {\SL Further Remarks on the Localization of Energy in the General Theory of Relativity}.
Ann. Phys., {\bf 12}, 118-133  (1961)

\bibitem{MTW} Misner,  C.W., Thorne,  K.S., Wheeler,  J.A.: {\SL Gravitation}, Freeman, San Francisco (1973)

\bibitem{MUELLER-1983} M\"uller-Hoissen, F., Nitsch, J.: {\SL Teleparallelism - A viable theory of gravity?}, 
  Phys. Rev. D {\bf 28}, 718  (1983)

\bibitem{MUELLER-1985} M\"uller-Hoissen, F., Nitsch, J.: {\SL On the Tetrad Theory of Gravity}, Gen. Rel. Grav. {\bf 17}, 747  (1985)

\bibitem{HEHL-1998} Muench, U., Gronwald, F., Hehl, F.W.: {\SL A small guide to variations in teleparallel gauge theories
of gravity and the Kaniel-Itin model}, {\tt http://arxiv.org/pdf/gr-qc/9801036}, (1998)

\bibitem{NESTER}  Nester, J.M.: {\SL Is there really a problem with the teleparallel theory?}, Class. Quant. Grav. {\bf 5} 1003-1010 (1988)

\bibitem{PENROSE} Penrose, R., Rindler, W.: {\SL Spinors and space-time}, Vol 1, Cambridge Univ. Press, Cambridge 1984 (1999)

\bibitem{PEREIRA} Pereira, J.G., Vargas, T., Zhang, C.M.: {\SL Axial-vector torsion and the teleparallel
Kerr spacetime}, Class. and Quant. Grav., {\bf 18}, 833-841 (2001)

\bibitem{SCHOUTEN} Schouten, J.: {\SL Ricci Calculus}, Springer, Berlin (1954)

\bibitem{SHIRAFUJI} Shirafuji, T., Nashed, G.G.L., Kobayashi, Y.:{\SL Equivalence Principle in the New General
Relativity}, {\tt http://arxiv.org/pdf/gr-qc/9609060} (1996)

\bibitem{STEPHANI} Stephani, H., {\SL Allgemeine Relativit\"atstheorie}, Deutscher Verlag der Wissenschaften, Berlin (1977)

\bibitem{Symbolic} K\"ohler, W.: ``Symbolic'', {\SL A script driven Java-program for symbolic tensor calculus in general relativity},
{\tt http://icgem.gfz-potsdam.de/Symbolic/Symbolic.html} (2010)

\bibitem{TUNG} Tung, R.S., Nester, J.M.: {\SL The quadratic spinor Lagrangian is equivalent to the teleparallel theory},
{\tt http://arxiv.org/pdf/gr-qc/9809030} (1999)

\bibitem{WIKI} Wikipedia, {\SL Alternatives to general relativity},
 {\tt http://en.wikipedia.org/wiki/Alternatives\_to\_general\_relativity}

\bibitem{WIKI-GAL-ROT} Wikipedia, {\SL Galaxy rotation curve},
{\tt http://en.wikipedia.org/wiki/Galaxy\_rotation\_curve}

\bibitem{WILL} Will,  C. M.: {\SL The Confrontation between General Relativity and Experiment}, 
Living Rev. Relat. {\bf 9} (3) {\tt http://livingreviews.org/lrr-2006-3}  (2006)

\bibitem{WILL-TEGP} Will,  C. M.: {\SL Theory and experiment in gravitational physics}, Revised Edition, 
Cambridge University Press, Cambridge (1993)

\end{thebibliography}
\end {document}